\pdfoutput=1

\documentclass[11pt]{article}

\usepackage[]{emnlp2021}

\usepackage{times}
\usepackage{latexsym}
\usepackage{microtype}
\usepackage{graphicx}
\usepackage{booktabs}  
\usepackage{multirow}
\usepackage{subfigure}
\usepackage{verbatim}
\usepackage{CJKutf8} 
\usepackage{amssymb}
\usepackage{amsmath}

\usepackage{algorithm}
\usepackage{algorithmic}
\usepackage{makecell}
\usepackage[T1]{fontenc}

\usepackage[utf8]{inputenc}

\usepackage{microtype}

\title{UserBERT: Contrastive User Model Pre-training}

\author{Chuhan Wu$^1$~~Fangzhao Wu$^2$~~Yang Yu$^3$~~Tao Qi$^1$~~Yongfeng Huang$^1$~~Xing Xie$^2$\\
    $^1$Department of Electronic Engineering \& BNRist, Tsinghua University, Beijing 100084, China  \\
     $^2$Microsoft Research Asia, Beijing 100080, China \\ 
     $^3$University of Science and Technology of China, Hefei 230027, China\\
  {\tt \{wuchuhan15, wufangzhao, taoqi.qt\}@gmail.com} \\
  {\tt tomyu613@icloud.com,} 
  {\tt yfhuang@tsinghua.edu.cn,}
  {\tt xingx@microsoft.com}
  }

\date{}

\begin{document}
\maketitle

\begin{abstract}

User modeling is critical for personalized web applications.
Existing user modeling methods usually train user models from user behaviors with task-specific labeled data.
However, labeled data in a target task may be insufficient for training accurate user models.
Fortunately, there are usually rich unlabeled user behavior data which encode rich information of user characteristics and interests.
Thus, pre-training user models on unlabeled user behavior data has the potential to improve user modeling for many downstream tasks.
In this paper, we propose a contrastive user model pre-training method named UserBERT.
Two self-supervision tasks are incorporated in UserBERT for user model pre-training on unlabeled user behavior data to empower user modeling. 
The first one is masked behavior prediction, which aims to model the relatedness between user behaviors.
The second one is behavior sequence matching, which aims to capture the inherent user interests that are consistent in different periods.
In addition, we propose a medium-hard negative sampling framework to select informative negative samples for better contrastive pre-training.
We maintain a synchronously updated candidate behavior pool and an asynchronously updated candidate behavior sequence pool to select the locally hardest negative behaviors and behavior sequences in an efficient way.
Extensive experiments on two real-world datasets in different tasks show that UserBERT can effectively improve  various user models.

\end{abstract}

\section{Introduction}

\begin{figure}[!t]
  \centering 
    \includegraphics[width=0.9\linewidth]{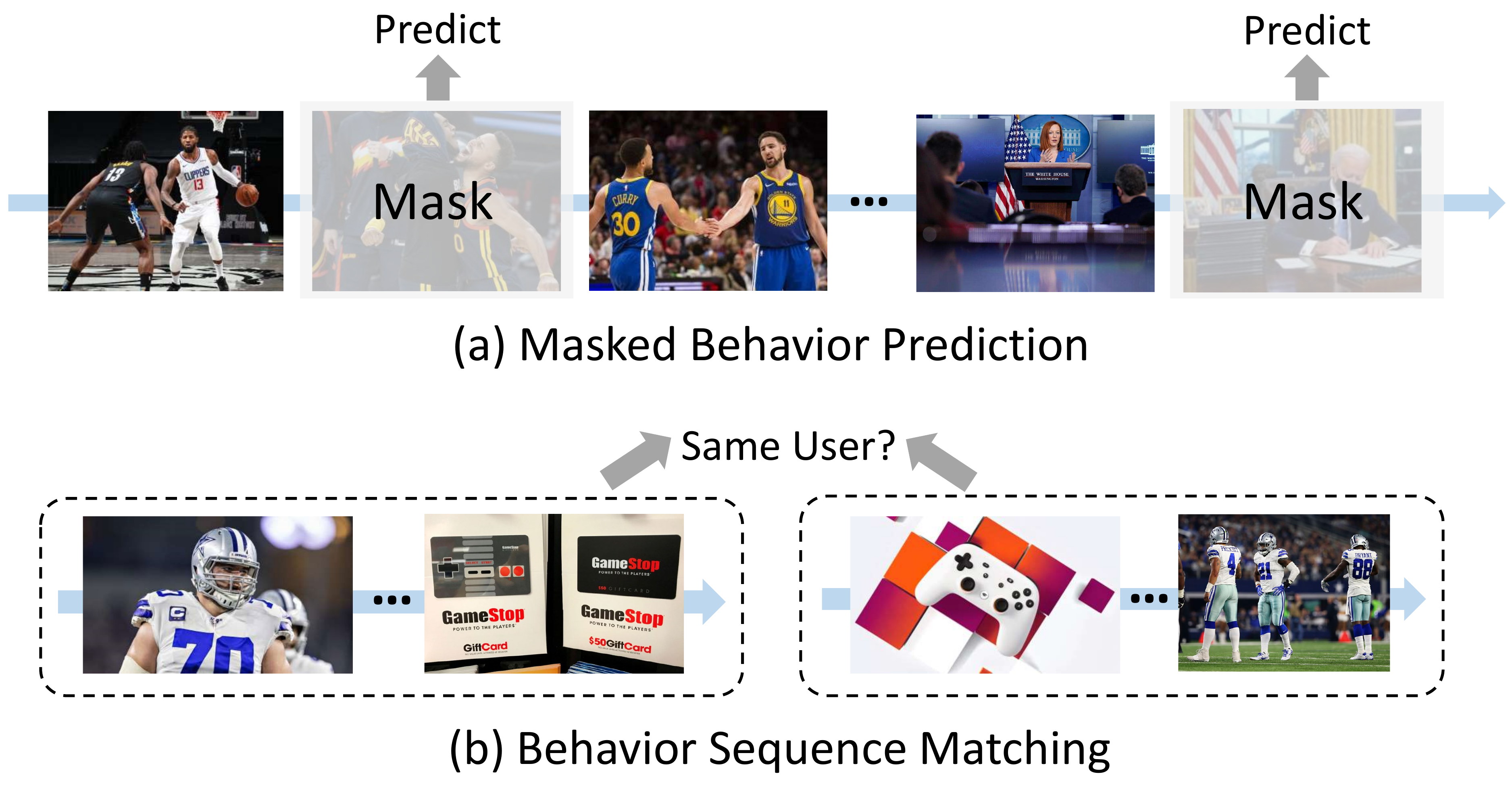} 
  \caption{Illustration of the two self-supervision tasks in UserBERT.}  \label{motivation}

\end{figure}

User modeling is a core technique for various  personalized web applications such as personalized news~\cite{okura2017embedding} and item recommendation~\cite{sun2019bert4rec}. 
Existing user modeling methods usually model users from their behaviors~\cite{zhou2019deep}.
For example,~\citet{wu2019neural} proposed a hierarchical user representation model for user demographic prediction that models users from their search queries.
\citet{an2019neural} proposed to use an attentive multi-view learning framework for native Ad click-through rate prediction, which models users based on their search query and webpage browsing behaviors.
These methods usually require a large amount of labeled data for training accurate user models~\cite{yuan2020parameter}.
However, in many scenarios labeled training data is insufficient, and it is often expensive and time-consuming to collect or annotate~\cite{wu2019neural}.
Fortunately, there are usually rich unlabeled user behaviors that encode rich information of user interests and characteristics~\cite{xie2020contrastive}.
For example, as shown in Fig.~\ref{motivation}(a), the news click behaviors can reveal the user's interests in sports and politics.
Thus, mining unlabeled user behavior data has the potential to generate a universal understanding of users to empower user modeling in downstream tasks~\cite{wu2020ptum}.

Motivated by the great success of pre-trained language models in NLP which are first pre-trained on large-scale unlabeled corpus via self-supervision and then finetuned in downstream NLP tasks, pre-training user models on unlabeled user behavior data may improve  user modeling for many user-intensive  tasks.
In many language model pre-training methods like BERT~\cite{devlin2019bert}, a masked token prediction task is used to capture the contexts of words.
In a similar way, predicting the masked user behavior in a user behavior sequence may help model the relations between user behaviors.
For example, as shown in Fig.~\ref{motivation}(a),
by predicting the masked user behaviors, the model can capture the relations between user behaviors and further help infer the user's interest (e.g., the Warriors team).
In addition, motivated by the sentence pair prediction task in language model pre-training for enhancing sentence modeling~\cite{vaswani2017attention,chi2020infoxlm}, matching the user behavior sequences may also be helpful for capturing user interests.
As shown in Fig.~\ref{motivation}(b), by matching the two behavior sequences from the same user in different periods, the model can better capture the inherent user interests (e.g., NFL) that is relatively stable over time.

In this paper, we propose a contrastive user model pre-training method named UserBERT.\footnote{We will soon release source codes and pre-trained models.}
In our approach, we pre-train the user models in two contrastive self-supervision tasks.
The first one is masked behavior prediction (MBP), which  is used to model  the relatedness between user behaviors.
The second one is behavior sequence matching (BSM).
It requires the model to identify whether two behavior sequences in different time periods come from the same user, which aims to capture the inherent user interests that are consistent over time.
Negative sampling is important for contrastive learning~\cite{kalantidis2020hard}.
However, random negative samples may not be informative while globally hardest negatives~\cite{xiong2020approximate} may be too confusing.
Thus, we propose a medium-hard negative sampling framework to select locally hardest samples for contrastive learning.
We first randomly construct a candidate behavior pool and a behavior sequence pool from the full candidate sets, and then select the top candidates that are most similar but not identical to the targets as negative samples.
We synchronously refresh the candidate behavior pool in each iteration, but asynchronously update the candidate behavior sequence pool after every certain number of iterations to reduce the computational cost.
Extensive experiments on two real-world datasets validate that our approach can consistently improve various user models and outperform several state-of-the-art user model pre-training methods.

The main contributions of this paper include: 
\begin{itemize}
    \item We propose a contrastive user model pre-training method that can exploit unlabeled user behaviors to pre-train user models.
    \item We propose a medium-hard negative sampling framework for more effective contrastive user model pre-training.
    \item We conduct extensive experiments on two real-world datasets to verify the effectiveness and advantage of our method.
\end{itemize}
\section{Related Work}

\subsection{User Model Pre-training}

In recent years, there are a few works on pre-training user models using unlabeled user behavior data via self-supervision~\cite{wu2020ptum,yuan2020parameter,xie2020contrastive}.
For example,~\citet{wu2020ptum} proposed a PTUM approach that uses a masked behavior prediction task and a next $K$ behaviors prediction tasks to pre-train user models.
\citet{yuan2020parameter} proposed a PeterRec approach that pre-trains user models in a masked behavior prediction task and an auto-regressive task that successively predicts user behaviors based on past ones.
However, user behaviors usually have much randomness, and only predicting specific user behaviors may not be optimal for modeling the inherent user characteristics.
\citet{xie2020contrastive} proposed a contrastive user model pre-training method that augments the user behavior sequences by the crop, mask and reorder operations, and matches the augmented behavior sequences from the same user.
However, in this method the augmented behavior sequences from the same users are usually highly overlapped, which may not be beneficial for the model to capture the inherent user interests.
In addition, the negative samples in all these methods are randomly drawn from the entire candidate sets, which usually have huge differences with the target positive samples.
These negative samples can be too easy for the model to distinguish, which may not be informative for  user model pre-training.
Different from these methods, our UserBERT approach incorporates a masked behavior prediction task to capture the relatedness between user behaviors and a behavior sequence matching task to capture the inherent user interests that are consistent in different periods.
In addition, our approach employs a medium-hard negative sampling framework to select locally hardest negative samples, which can help learn more discriminative user models.

\subsection{Negative Sampling for Contrastive Learning}

Negative sampling techniques for contrastive learning have been extensively studied~\cite{chen2020simple,robinson2020contrastive}.
For example,~\citet{oord2018representation} used  randomly selected negative samples for contrastive learning.
\citet{xiao2017joint} proposed to use a circular queue to store the features of these negative candidates that are included in recent batches.
\citet{he2020momentum} proposed to maintain a negative candidate pool of recent batches and update their representations with gradient momentum. 
In these methods negative candidates are in fact uniformly selected from the full candidate set, which usually have significant differences with the positive samples and may be too easy for the model to discriminate.
Researchers found that many of these negatives are not informative for model training~\cite{kalantidis2020hard}.
Thus, several methods explore to select hard negatives for training discriminative models~\cite{xiong2020approximate,robinson2020contrastive}.
For example,~\citet{xiong2020approximate} proposed to use an asynchronously updated ANN index to select the globally hardest negative samples. 
However, globally hardest samples may be too confusing for the model and can sometimes be misleading, which may be suboptimal.
In our approach, we propose a medium-hard negative sampling method to select locally hardest negatives, which may be more appropriate for learning an accurate and discriminative model.

\section{UserBERT}\label{sec:Model}

\subsection{General User Model Framework}

\begin{figure}[!t]
  \centering
    \includegraphics[width=0.32\textwidth]{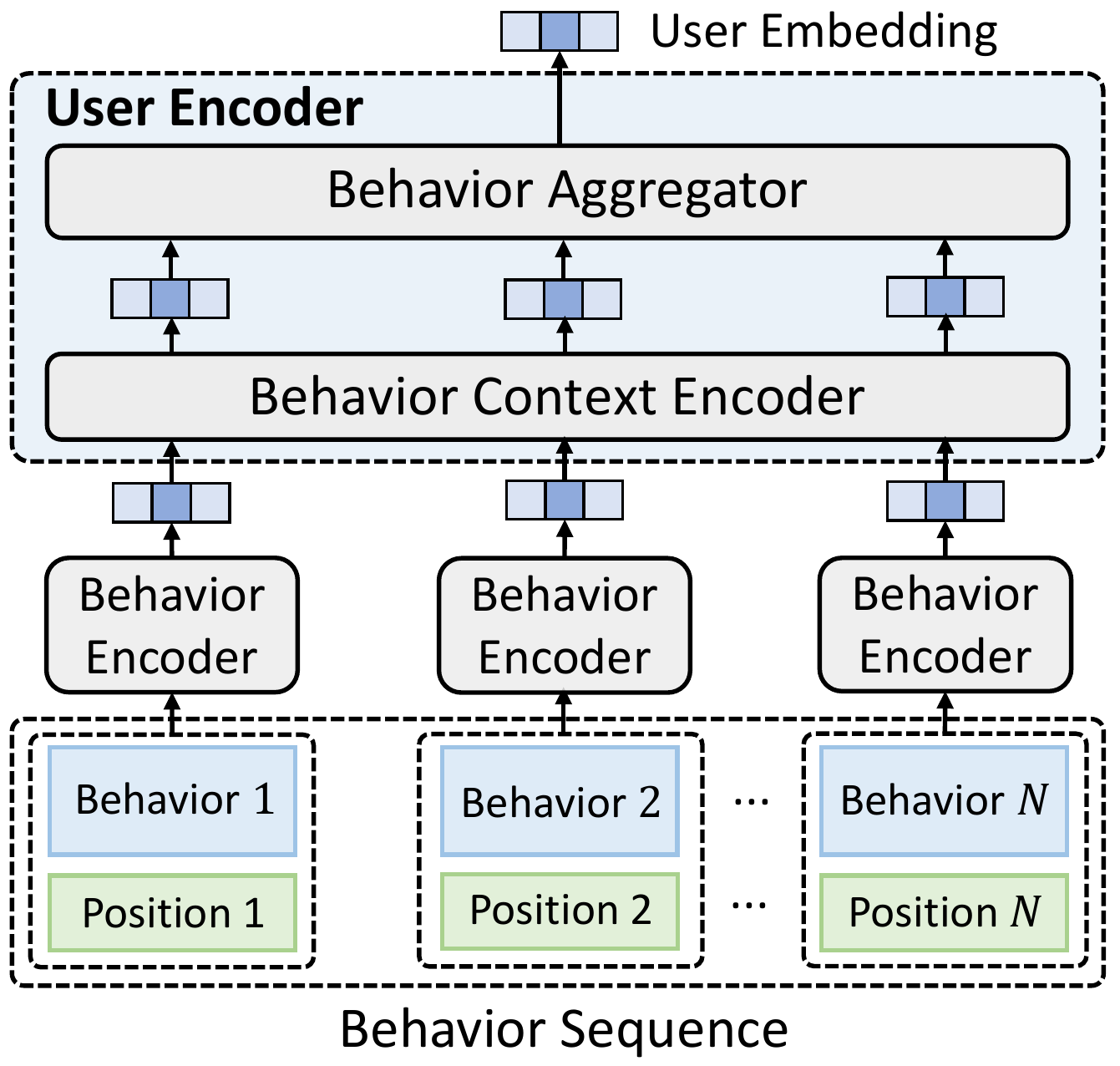}
  \caption{A general framework of the user model.}  \label{um}
\end{figure}

We first briefly introduce the general framework of  user model, as shown in Fig.~\ref{um}.
It leverages a hierarchical architecture with a behavior encoder and a user encoder.
The behavior encoder transforms each user behavior and its position in the behavior sequence into its embedding.
It can be implemented by various models according to the characteristics of user behaviors.
For example, for ID-based user behaviors, it can be implemented with an ID embedding table~\cite{sun2019bert4rec}.
For user behaviors that involve textual information, it can be neural NLP models like CNN~\cite{kim2014convolutional} and Transformer~\cite{vaswani2017attention} networks. 
The user encoder takes the behavior embedding sequence as input and encodes it into a user embedding.
It contains two submodules, i.e., a behavior context encoder to learn hidden behavior representations by capturing the contexts of behaviors, and a behavior aggregator that summarizes the contextual behavior representations into a unified user embedding.
The behavior context encoder can be implemented by many  models like CNN~\cite{wu2019neural}, LSTM~\cite{an2019neural} and self-attention~\cite{wu2019nrms} networks, and the behavior aggregator can be a max pooling, average pooling, attentive pooling~\cite{yang2016hierarchical} or last pooling module~\cite{hidasi2016gru}.
By pre-training the user model with unlabeled user behaviors via self-supervision, the model can exploit the universal user information conveyed by user behaviors to empower downstream tasks.

\begin{figure*}[!t]
  \centering
  \subfigure[Masked Behavior Prediction (MBP) task.]{
    \includegraphics[height=1.5in]{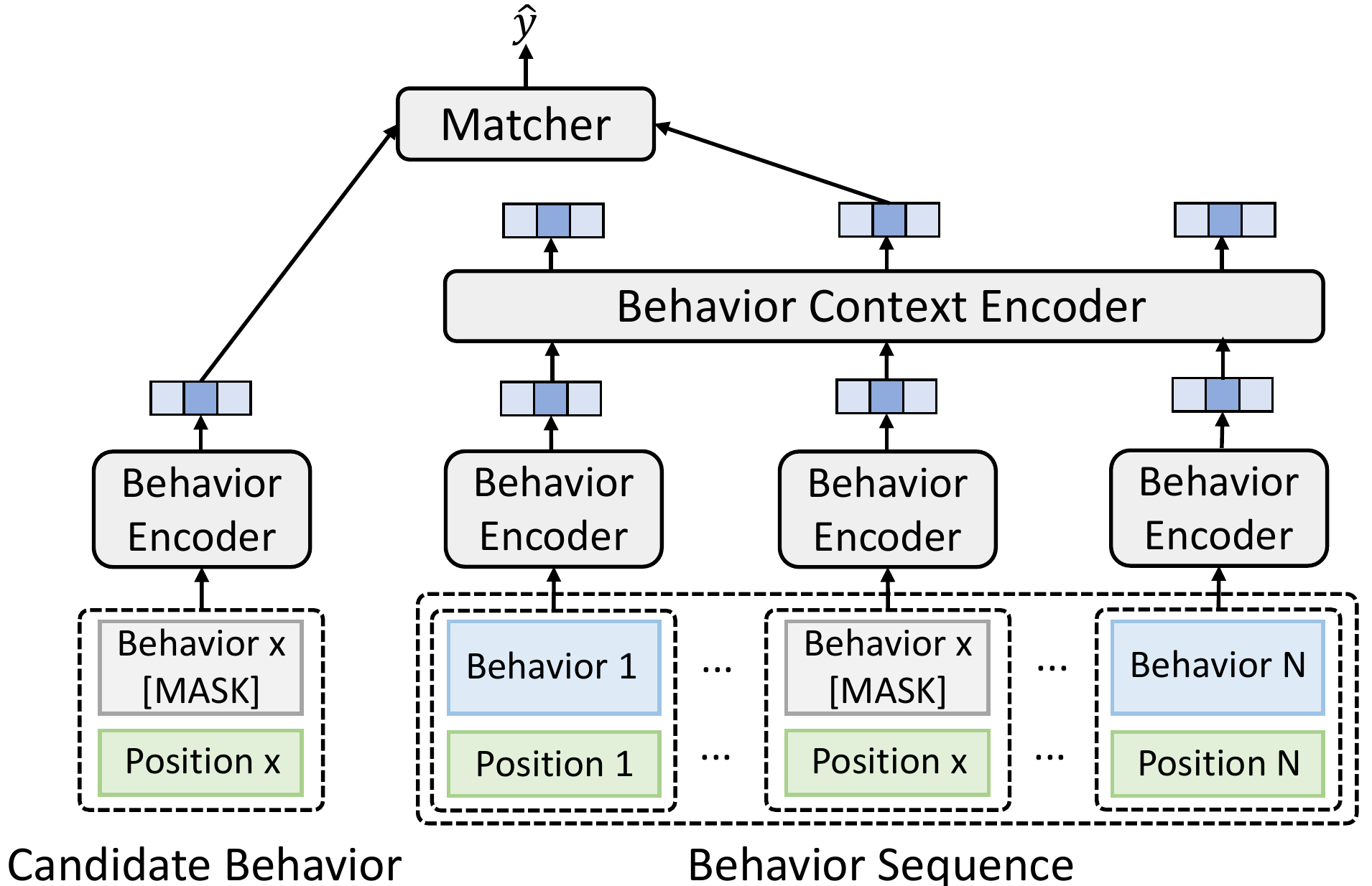}\label{task1}
    }\hspace{0.1in}
      \subfigure[Behavior Sequence Matching (BSM) task.]{
    \includegraphics[height=1.61in]{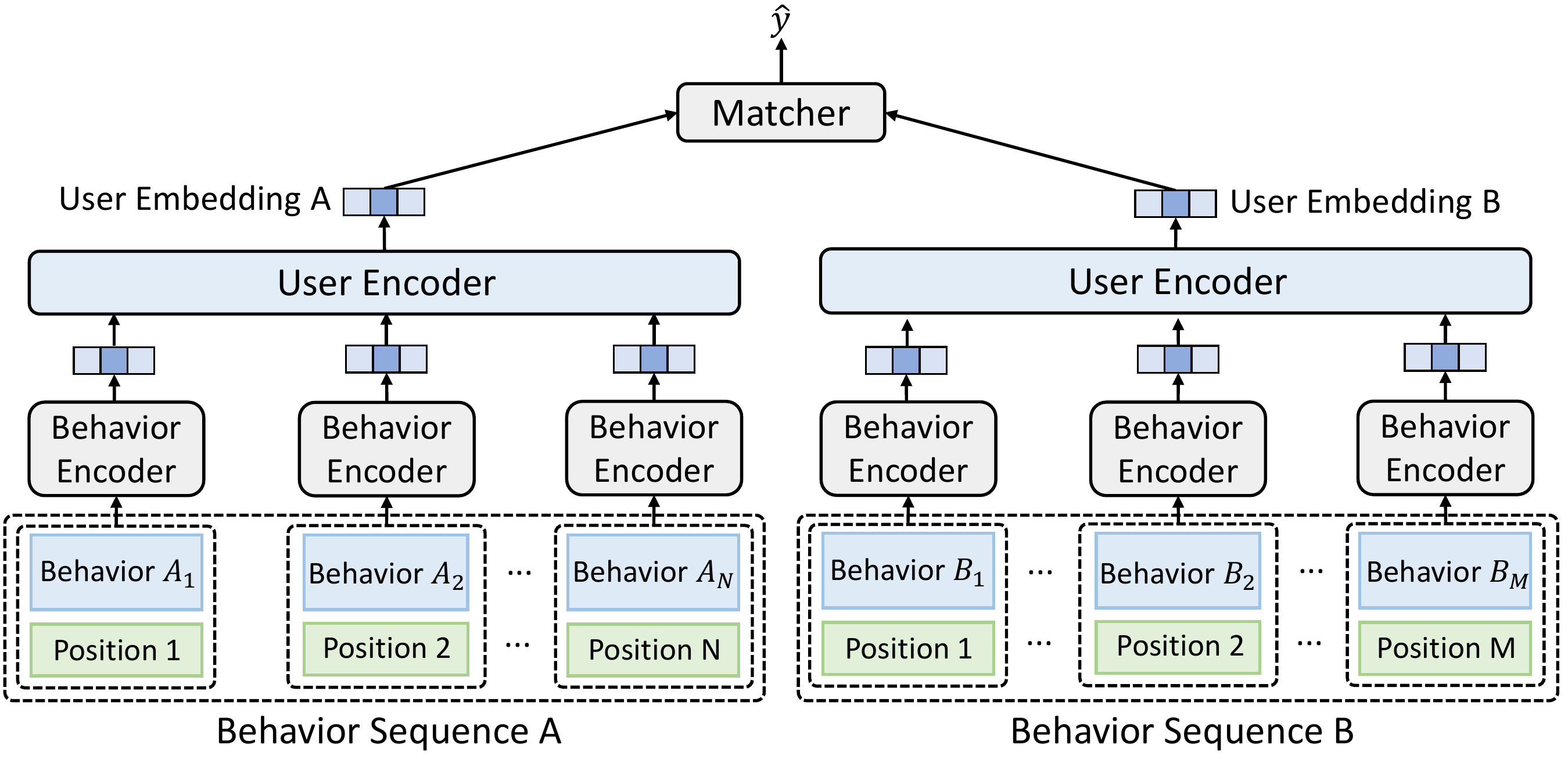}\label{task2}
    }
  \caption{Frameworks of the two self-supervision tasks for user model pre-training.}
\end{figure*}

\subsection{Contrastive Pre-training Tasks}

Next, we introduce the two self-supervision tasks for contrastive user model pre-training, i.e., masked behavior prediction (MBP) and behavior sequence matching (BSM).

\subsubsection{Masked Behavior Prediction (MBP)}

Following PTUM~\cite{wu2020ptum}, we use a masked behavior prediction task with small variants to capture the relations between behaviors of the same user, as shown in Fig.~\ref{task1}.
In PTUM the masked behavior is predicted from the user embedding that does not encode masked positions.
Instead, we use the hidden representations (generated by the behavior context encoder) at the masked positions to predict the masked behavior to better capture masked behaviors' positions.
We randomly mask 10\% of user behaviors (and at least one) in a user behavior sequence, and pre-train the model in a contrastive way.
For each masked behavior $x$ (regarded as the positive candidate) we sample $K$ negative candidate behaviors, and we use a matcher to jointly predict the matching scores of these $K+1$ candidates based on their relevance to the hidden representation at the masking position.
These scores are further normalized by softmax, and the task is formulated as a $K+1$-way classification problem.
The loss function we used is cross-entropy, which is formulated as follows:
\begin{equation}
    \mathcal{L}_{MBP}=-\sum_{x\in \mathcal{M}}\sum_{i=1}^{K+1}{y^x_i \log(\hat{y}^x_i)},
\end{equation}
where $y^x_i$ and $\hat{y}^x_i$ are gold and predicted labels of the $i$-th candidate behavior given the masked behavior $x$, and $\mathcal{M}$ is the set of masked behaviors.

\subsubsection{Behavior Sequence Matching (BSM)}

The second pre-training task is behavior sequence matching (BSM).
As shown in Fig.~\ref{task2}, the goal of this task is to identify whether two behavior sequences A and B come from the same user.
To enforce the model to capture the inherent user interests that are relatively stable across different time periods, we ensure that the behavior sequences A and B have no overlap in time.
We use the user model to encode both sequences and evaluate their similarity via a matcher.
For each pair of behavior sequences from the same user (sequence B is regarded as the positive candidate), we sample $P$ negative candidate behavior sequences from other users that do not overlap with sequence A in time.
We jointly predict the matching scores of the $P+1$ behavior sequences and normalize them via softmax.
The loss function of the behavior sequence matching task is also cross-entropy, which is formulated as follows:
\begin{equation}
    \mathcal{L}_{BSM}=-\sum_{A\in \mathcal{S}}\sum_{i=1}^{P+1}{y^A_i \log(\hat{y}^A_i)},
\end{equation}
where $y^A_i$ and $\hat{y}^A_i$ are the gold label and predicted score of the $i$-th candidate behavior sequence given the target sequence $A$, and $\mathcal{S}$ is the set of user behavior sequences for model pre-training.
We jointly pre-train the user model in both MBP and BSM tasks, and the unified loss function $\mathcal{L}$ is a summation of the two loss functions as follows:
\begin{equation}
    \mathcal{L} =\mathcal{L}_{MBP} + \mathcal{L}_{BSM}.
\end{equation}

\begin{figure*}[!t]
  \centering 
    \includegraphics[width=0.91\textwidth]{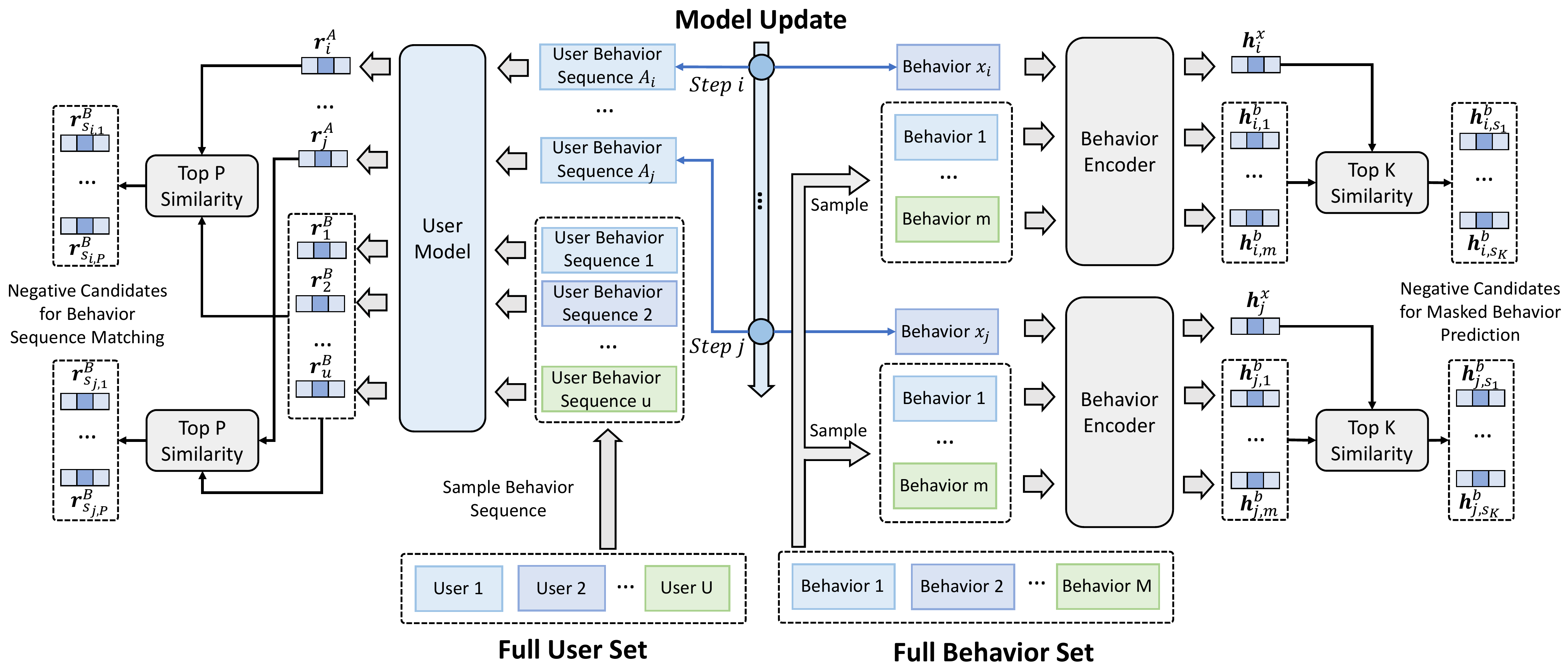} 
  \caption{The framework of our medium-hard negative sampling method.}  \label{sample}

\end{figure*}

\subsection{Medium-Hard Negative Sampling}

Finally, we introduce our proposed medium-hard negative sampling method to select the locally hardest negative candidate behaviors and candidate behavior sequences for contrastive pre-training.
As shown in Fig.~\ref{sample}, it contains two modules, i.e., negative behavior sampling (right) and negative user  behavior sequence sampling (left).
The negative behavior sampling module aims to select negative candidates for the masked behavior prediction task.
Negatives uniformly selected from the full behavior set may not be informative because they are usually easy to be distinguished.
However, the globally hardest negatives may be too puzzling or even misleading to the model.
For example, the hardest negative for a target webpage browsing behavior of ``NFL Schedule: Week 1'' can be a webpage entitled ``NFL Schedule | Week 1'', which has no substantial difference with the target.
To handle this problem, we propose to construct a smaller candidate behavior pool and select the locally hardest samples from this pool as ``medium-hard'' negatives.
We assume that the entire behavior set $\mathcal{B}$ contains $M$ user behaviors, and we randomly sample $m$ ($m<<M$) behaviors to construct a candidate behavior pool. 
This pool is synchronously updated during the model training, i.e., re-sampling user behaviors from $\mathcal{B}$ at each model training step.
We denote a masked behavior at the $i$-th model update step as $x_i$ and the $m$ behaviors in the corresponding candidate behavior pool as $[b_1, ..., b_m]$.
We use the behavior encoder in the user model to encode these behaviors into their hidden representations, which are respectively denoted as $\mathbf{h}^x_i$ and $\mathbf{H}^b=[\mathbf{h}^b_1, ..., \mathbf{h}^b_m]$.
We evaluate the cosine similarities between $\mathbf{h}^x_i$ and each behavior embedding in $\mathbf{H}^b$, and select $K$ behaviors with top similarity scores as negative candidates for masked behavior prediction (candidates that are identical to the target are filtered).
In this way, the negatives are harder if $m$ is larger.
We can adjust the difficulties of negative samples by tuning the value of $m$.

The negative user  behavior sequence sampling module aims to select negative candidates for the behavior sequence matching task, which uses a similar way to sample medium-hard negatives by choosing the locally hardest samples in a candidate behavior sequence pool.
However, different from processing user behaviors, encoding  behavior sequences is much more time-consuming.
Thus, we construct a behavior sequence pool that is asynchronously updated after every certain number of iterations to reduce the computational cost.
We denote the target user behavior sequence from the $i$-th to the $j$-th training steps as $[A_i, ..., A_j]$.
We randomly crop $u$ behavior sequences from the behavior sequences of other users in the full user set to form the candidate  behavior sequence pool, which is denoted as $[B_1, ..., B_u]$.
We ensure that these sequences do not have overlaps with $[A_i, ..., A_j]$ in time.
This pool is static between the $i$-th and $j$-th training steps (the interval is $j-i+1$ steps).
We use the user model at the $i$-th training step to encode target and candidate behavior sequences, which are respectively denoted as $\mathbf{R}^A=[\mathbf{r}^A_i, ..., \mathbf{r}^A_j]$ and $\mathbf{R}^B=[\mathbf{r}^B_1, ..., \mathbf{r}^B_u]$.
We use cosine distance to measure the similarity between each target behavior sequence embedding in $\mathbf{R}^A$ and each candidate in $\mathbf{R}^B$, and select the candidates with the highest similarities to the target sequence as negative samples.\footnote{Need to re-encode the target and candidate behavior sequences using the currently updated user model for training.}
In this way, we can obtain medium-hard candidate user behavior sequences efficiently by using a moderate pool size and update interval.

\begin{table}[!h]
\centering
\resizebox{0.48\textwidth}{!}{
\begin{tabular}{lrlr}
\Xhline{1.5pt}
\multicolumn{4}{c}{\textbf{Demo}}                                                \\ \hline
\# Users                  & 10,000 & \# News                        & 42,255      \\
\# Impressions            & 360,428 & \# Clicked samples            & 503,698    \\  \hline
\multicolumn{4}{c}{\textbf{CTR}}                                                 \\ \hline
\# Users                  & 374,584 & \# Ads                        & 4,159      \\
\# Impressions            & 400,000 & \# Clicked samples            & 364,281    \\ \hline
\# Users for pre-training & 500,000 & \# Behaviors for pre-training & 63,178,293 \\ \Xhline{1.5pt}
\end{tabular}
}   
\caption{Statistics of the two datasets.}\label{dataset} \vspace{0.0in}
\end{table}

\begin{table*}[!t]
	\centering
\resizebox{0.93\linewidth}{!}{
\begin{tabular}{lcccccc}
\Xhline{1.5pt} 
\multirow{2}{*}{\textbf{Methods}} & \multicolumn{2}{c}{\textbf{10\%}} & \multicolumn{2}{c}{\textbf{25\%}} & \multicolumn{2}{c}{\textbf{100\%}} \\ \cline{2-7} 
                                  & AUC             & nDCG@10              & AUC             & nDCG@10              & AUC              & nDCG@10              \\ \hline
NAML           & 58.83$\pm$0.27     & 37.54$\pm$0.24     & 60.20$\pm$0.23     & 38.25$\pm$0.22     & 61.02$\pm$0.20   & 38.84$\pm$0.19  \\
NAML+CP        & 59.68$\pm$0.25     & 37.96$\pm$0.23     & 60.90$\pm$0.21     & 38.94$\pm$0.20     & 61.59$\pm$0.18   & 39.40$\pm$0.17  \\
NAML+PTUM      & 60.22$\pm$0.23     & 38.27$\pm$0.24     & 61.53$\pm$0.20     & 39.33$\pm$0.19     & 61.94$\pm$0.17   & 39.78$\pm$0.18  \\
NAML+UserBERT  & 60.85$\pm$0.24     & 38.81$\pm$0.24     & 61.94$\pm$0.19     & 39.76$\pm$0.21     & 62.56$\pm$0.16   & 40.30$\pm$0.18  \\ \hline
LSTUR          & 59.31$\pm$0.25     & 37.77$\pm$0.25     & 60.60$\pm$0.20     & 38.66$\pm$0.19     & 61.67$\pm$0.18   & 39.49$\pm$0.19       \\
LSTUR+CP       & 60.12$\pm$0.24     & 38.18$\pm$0.23     & 61.28$\pm$0.18     & 39.16$\pm$0.17     & 62.05$\pm$0.16   & 39.81$\pm$0.18     \\
LSTUR+PTUM     & 60.61$\pm$0.22     & 38.65$\pm$0.22     & 61.75$\pm$0.17     & 39.59$\pm$0.18     & 62.48$\pm$0.15   & 40.21$\pm$0.16   \\
LSTUR+UserBERT & 61.27$\pm$0.23     & 39.13$\pm$0.20     & 62.34$\pm$0.18     & 40.13$\pm$0.18     & 62.95$\pm$0.16   & 40.73$\pm$0.15    \\ \hline
NRMS           & 59.22$\pm$0.20     & 37.70$\pm$0.21     & 60.53$\pm$0.18     & 38.59$\pm$0.19     & 61.58$\pm$0.16   & 39.38$\pm$0.14      \\
NRMS+CP        & 60.05$\pm$0.21     & 37.99$\pm$0.20     & 61.26$\pm$0.17     & 39.13$\pm$0.18     & 61.95$\pm$0.15   & 39.80$\pm$0.15    \\
NRMS+PTUM      & 60.58$\pm$0.18     & 38.62$\pm$0.18     & 61.71$\pm$0.16     & 39.57$\pm$0.15     & 62.32$\pm$0.13   & 40.10$\pm$0.15   \\
NRMS+UserBERT  & \textbf{61.24}$\pm$0.16     & \textbf{39.10}$\pm$0.17     & \textbf{62.33}$\pm$0.15     & \textbf{40.10}$\pm$0.14     & \textbf{62.87}$\pm$0.14   & \textbf{40.64}$\pm$0.12            \\  
\Xhline{1.5pt}
\end{tabular}
}

\caption{Results on \textit{News} dataset with different ratios of training data.}\label{table.result}
\vspace{0.0in}
\end{table*}

\begin{table*}[!t]
	\centering
\resizebox{0.93\linewidth}{!}{
\begin{tabular}{lcccccc}
\Xhline{1.5pt} 
\multirow{2}{*}{\textbf{Methods}} & \multicolumn{2}{c}{\textbf{10\%}} & \multicolumn{2}{c}{\textbf{25\%}} & \multicolumn{2}{c}{\textbf{100\%}} \\ \cline{2-7} 
                            & AUC       & AP              & AUC             & AP              & AUC              & AP              \\ \hline
GRU4Rec                           & 70.96$\pm$0.11           & 72.79$\pm$0.10           & 71.51$\pm$0.09           & 73.26$\pm$0.07           & 72.20$\pm$0.06            & 74.40$\pm$0.07           \\
GRU4Rec+CP                        & 71.45$\pm$0.12           & 73.62$\pm$0.12           & 71.94$\pm$0.10           & 74.13$\pm$0.10           & 72.68$\pm$0.09            & 75.23$\pm$0.08           \\
GRU4Rec+PTUM                      & 71.94$\pm$0.13  & 74.18$\pm$0.14  & 72.37$\pm$0.12 & 74.59$\pm$0.11 & 72.79$\pm$0.10  & 75.40$\pm$0.10   \\
GRU4Rec+UserBERT                  & 72.67$\pm$0.10  & 74.75$\pm$0.10  & 72.90$\pm$0.11 & 75.01$\pm$0.11 & 73.24$\pm$0.09  & 75.89$\pm$0.09   \\ \hline
NativeCTR                         & 71.13$\pm$0.10  & 72.95$\pm$0.09  & 71.69$\pm$0.08 & 73.51$\pm$0.07 & 72.35$\pm$0.08  & 74.56$\pm$0.06   \\
NativeCTR+CP                      & 71.81$\pm$0.09  & 73.69$\pm$0.09  & 72.20$\pm$0.08 & 74.50$\pm$0.09 & 72.77$\pm$0.07  & 75,40$\pm$0.07   \\
NativeCTR+PTUM                    & 72.19$\pm$0.09  & 74.14$\pm$0.08  & 72.58$\pm$0.07 & 74.85$\pm$0.07 & 72.91$\pm$0.06  & 75.57$\pm$0.06   \\
NativeCTR+UserBERT                & 72.91$\pm$0.08  & 74.92$\pm$0.07  & 73.14$\pm$0.08 & 75.20$\pm$0.08 & 73.40$\pm$0.07  & 76.09$\pm$0.07   \\ \hline
BERT4Rec                          & 71.25$\pm$0.10  & 73.02$\pm$0.09  & 71.89$\pm$0.07 & 74.12$\pm$0.06 & 72.99$\pm$0.06  & 75.45$\pm$0.05   \\
BERT4Rec+CP                       & 72.03$\pm$0.08  & 74.11$\pm$0.09  & 72.65$\pm$0.06 & 75.24$\pm$0.07 & 73.39$\pm$0.06  & 76.12$\pm$0.05   \\
BERT4Rec+PTUM                     & 72.30$\pm$0.08  & 74.39$\pm$0.07  & 72.89$\pm$0.07 & 75.44$\pm$0.06 & 73.59$\pm$0.05  & 76.48$\pm$0.05   \\
BERT4Rec+UserBERT                 & \textbf{73.12}$\pm$0.09 & \textbf{75.04}$\pm$0.09 & \textbf{73.39}$\pm$0.07 & \textbf{75.34}$\pm$0.06 & \textbf{73.96}$\pm$0.06            & \textbf{76.72}$\pm$0.06           \\  
\Xhline{1.5pt}
\end{tabular}
}

\caption{Results on \textit{CTR} dataset with different ratios of training data.}\label{table.result2}
\vspace{0.0in}
\end{table*}

\section{Experiments}\label{sec:Experiments}

\subsection{Datasets and Experimental Settings}
In our experiments, we conduct experiments in two tasks.
The first task is news recommendation.
We take the dataset (denoted as \textit{News}) used in~\cite{wu2019nrhub}.
It contains news click behaviors of 10,000 users on Microsoft News and webpages browsing behaviors on Bing in one month.
As studied in~\cite{wu2019nrhub}, the goal is to predict future news clicks of users based on the titles of their browsed webpages.
The second task is Ad CTR prediction.
The dataset (denoted as \textit{CTR}) contains Ad titles and descriptions, Ad impressions, and the webpage browsing behaviors of 374,584 users in one month, which is  used by~\cite{wu2020ptum}.
The goal is to predict whether a user clicks a candidate Ad based on Ad texts and the titles of this user's browsed webpages.
In the two tasks, we used the same dataset split strategy as prior works~\cite{wu2019nrhub,wu2020ptum}
We used the same unlabeled user behavior dataset for user model pre-training, which contains the titles of browsed webpages of 500,000 users on Bing in six months.
The dataset statistics are summarized in Table~\ref{dataset}.

In our experiments, for fair comparison we followed the base user models settings in PTUM~\cite{wu2020ptum}.
We used dot product to implement the matchers in our approach.
The number of negative candidates was 4.
The sizes of candidate behavior and behavior sequence pools were 1,000 and 100, respectively.
The behavior sequence pool update interval was 50 steps.
We used Adam~\cite{kingma2014adam} as the optimizer.
The learning rate was 1e-4.
Detailed hyperparameter configurations are included in supplementary materials.
These hyperparameters were tuned according to the validation performance.
We used AUC and nDCG@10 to measure model performance on the \textit{News} dataset, and used AUC and AP as metrics on the \textit{CTR} dataset.
We repeated each experiment 5 times and reported the average results.

\subsection{Performance Evaluation}

In this section, we compare the performance of our approach with several baseline methods, including (a) w/o pre-training; (b) PTUM~\cite{wu2020ptum} and (c) the contrastive pre-training method (denoted as CP) proposed in~\cite{xie2020contrastive}.
On the \textit{News} dataset we use NAML~\cite{wu2019},  LSTUR~\cite{an2019neural}, and NRMS~\cite{wu2019nrms}, which are widely used news recommendation methods~\cite{wu2020mind}.
on the \textit{CTR} dataset we use 
 GRU4Rec~\cite{hidasi2016gru}, NativeCTR~\cite{an2019native} and BERT4Rec~\cite{sun2019bert4rec} to implement the user models.
The performance under different percentage of training data on the two datasets is respectively shown in Tables~\ref{table.result} and \ref{table.result2}, which leads to several findings.
First, pre-trained user models consistently achieve better performance than the models without pre-training, and the advantage is larger when less labeled data is used for training.
This is because pre-trained user models can exploit universal user information encoded by unlabeled user behaviors to enhance user modeling, which can reduce the dependency of user models on labeled data.
Second,  PTUM and UserBERT achieve better performance than CP.
This may be because in the CP method, the behavior sequences augmented from the same user may have many overlaps, which is not beneficial for the model to capture the inherent user interests.
Third, UserBERT consistently outperforms PTUM and the improvement is significant ($p<0.05$ in t-tests).
This is probably because PTUM uses self-supervision tasks that only predict specific behaviors, which may be disturbed by the randomness of user behaviors.
Our UserBERT approach uses a behavior sequence matching task to capture the intrinsic user interests, which may be more robust to the noisy behaviors.
In addition, in PTUM negative samples are randomly selected, which may not be very informative nor representative.
In contrast, UserBERT is trained with more informative negatives with a medium-hard negative sampling method, which yields better performance.

\begin{figure}[!t]
	\centering
	\includegraphics[width=0.4\textwidth]{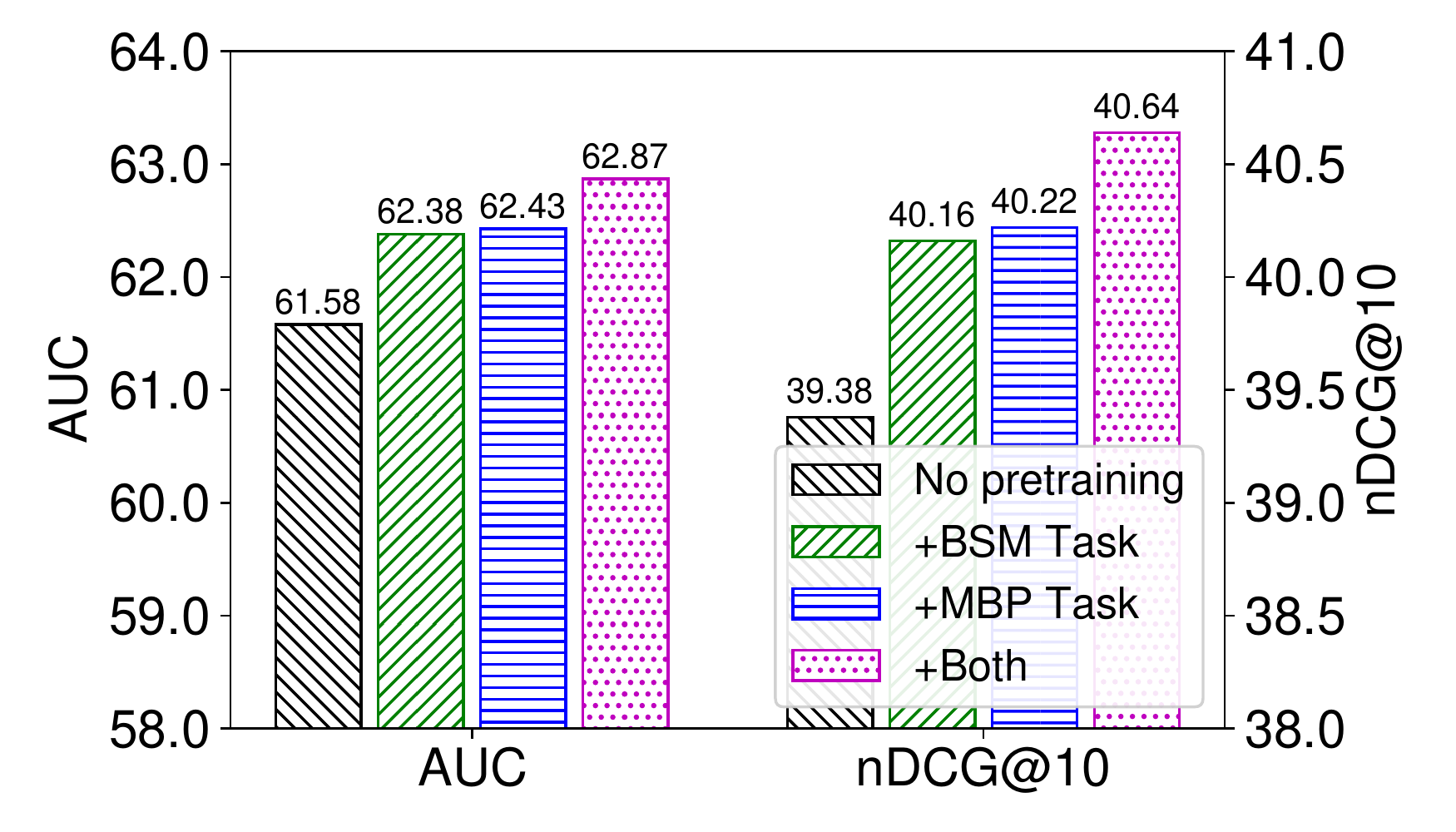}

\vspace{0.0in}
\caption{Effect of the two pre-training tasks.}\label{fig.task}  
\vspace{0.0in}
\end{figure}

\begin{figure}[!t]
	\centering

	\includegraphics[width=0.4\textwidth]{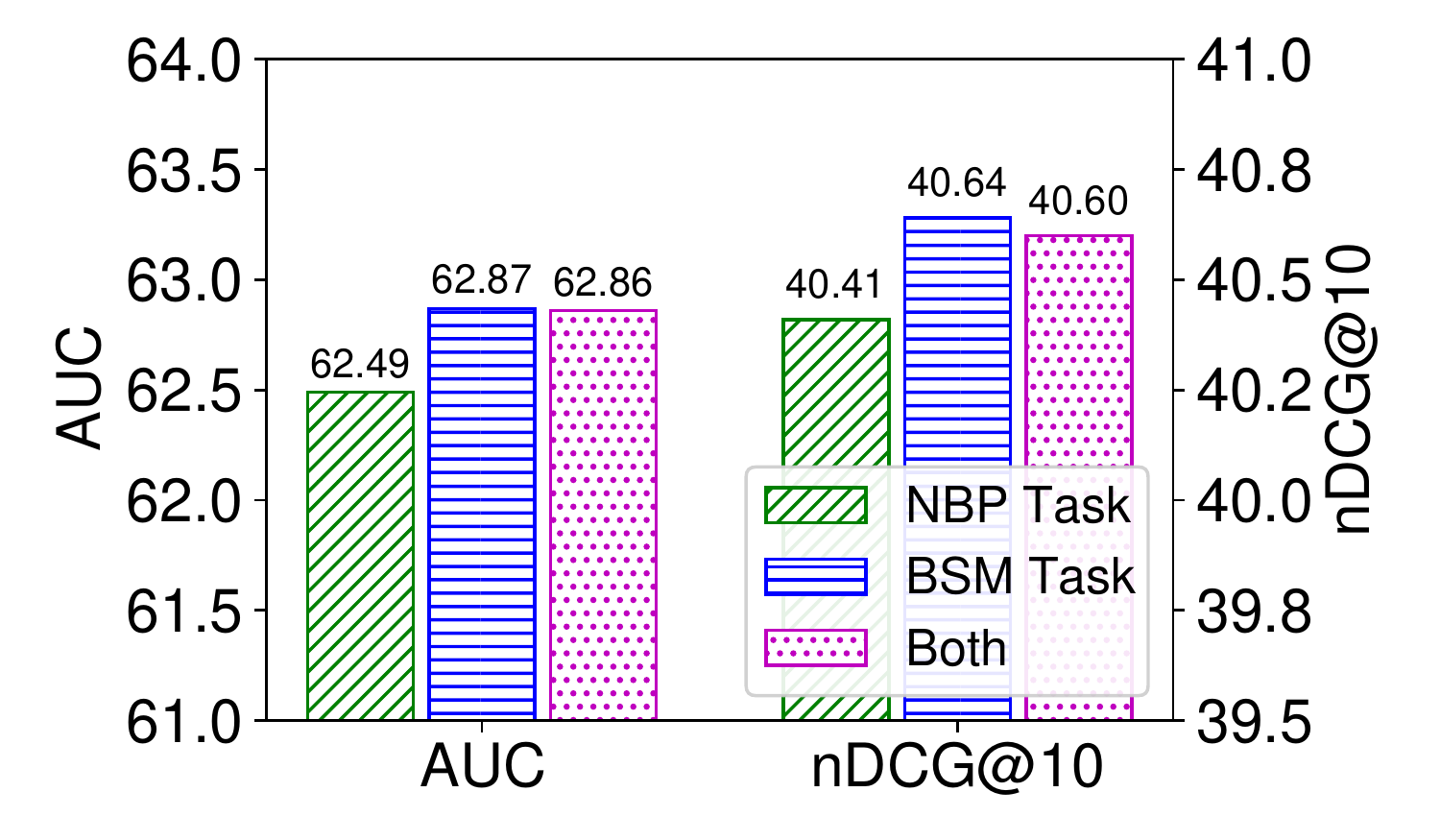}
	
	\vspace{0.0in}
\caption{Comparisons of BSM task and NBP task.}\label{fig.task2} 
\vspace{0.0in} 
\end{figure}

\begin{figure}[!t]
	\centering

	\includegraphics[width=0.4\textwidth]{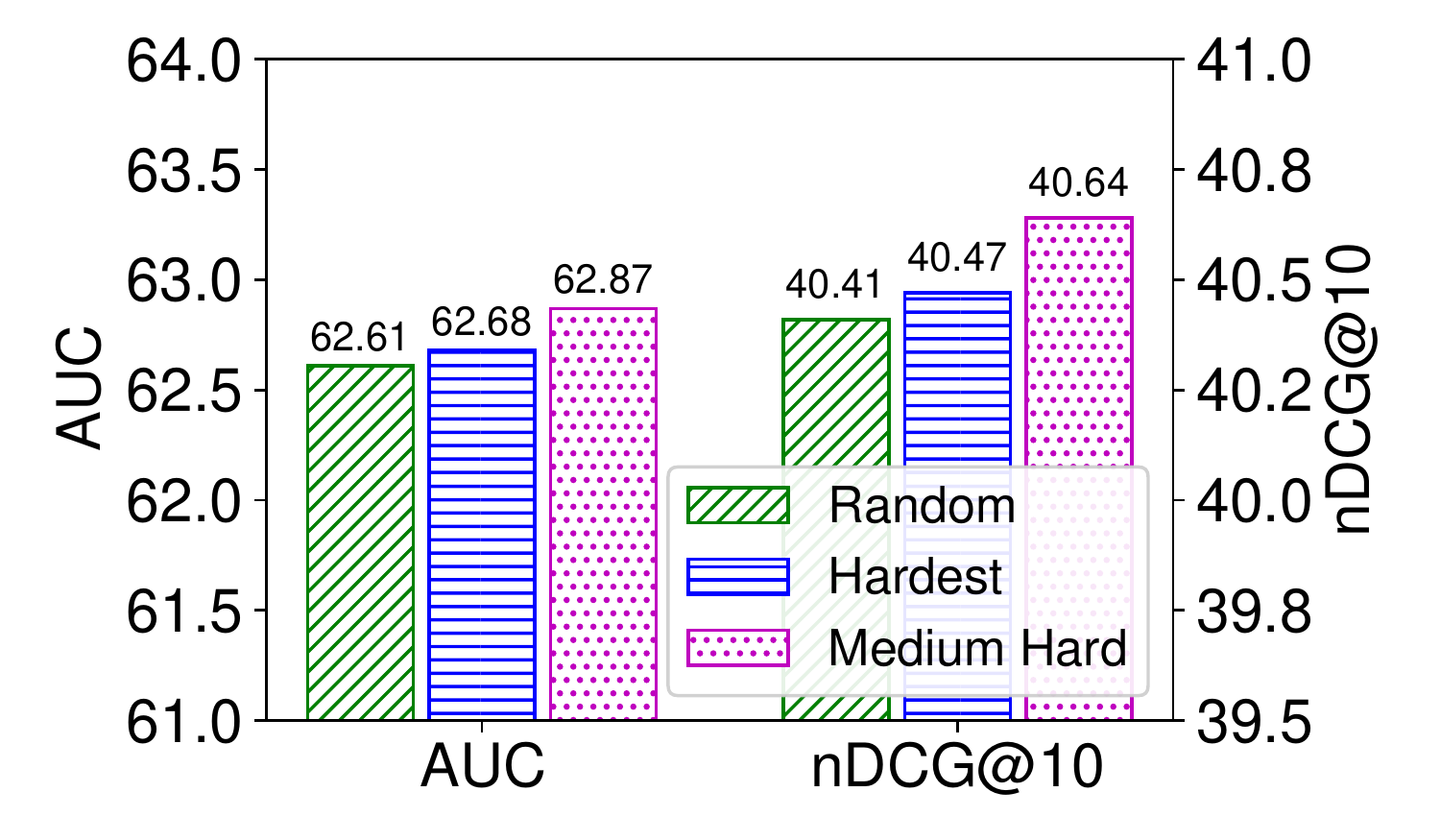}

	\vspace{0.0in}
\caption{Effect of medium-hard negative sampling.}\label{fig.sample}  
\vspace{0.0in}
\end{figure}

\subsection{Influence of Pre-training Tasks}

In this section, we study the influence of the self-supervision tasks for user model pre-training.
We first verify the effectiveness of each task used in UserBERT.
We compare the results of NRMS on the \textit{News} dataset and BERT4Rec on the \textit{CTR} dataset without pre-training or pre-trained with different tasks. \footnote{Same base user models are used in the rest experiments.}
Due to space limitation, we only present the results on the \textit{News} dataset in the main content, and the results on the \textit{CTR} dataset are in supplements.
From the results shown in Fig.~\ref{fig.task}, we find that both BSM and MBP tasks are very useful for model pre-training.
This may be because the MBP task can encourage the user model to capture the relatedness between behaviors and the BSM task helps to model the inherent user interests that are consistent across different time periods.
In addition, combining both tasks is better than using a single one.
It shows that the two tasks can provide complementary information for each other to improve user model pre-training.

Then, we compare our newly proposed BSM task with the next $K$ behaviors prediction (NBP) task from  PTUM.
We evaluate the model performance using one of the NBP and BSM tasks or combining both of them (the MBP task is still preserved).
The results are shown in Fig.~\ref{fig.task2}.
We find the models pre-trained in the BSM task outperforms those pre-trained in the NBP task.
This may be because the NBP task in fact has some overlaps with the MBP task (can be regarded as masking the last $K$ behaviors), and predicting the exact behaviors will also be disturbed by the behavior randomness.
In contrast, the BSM task can also help the model capture the relatedness between past and future behaviors and may be more robust to the noisy behaviors.
In addition, we do not observe any performance gain when combining the BSM and NBP tasks.
Thus, we only use the MBP and BSM tasks in our approach.

\begin{figure}[!t]
	\centering
	\includegraphics[width=0.36\textwidth]{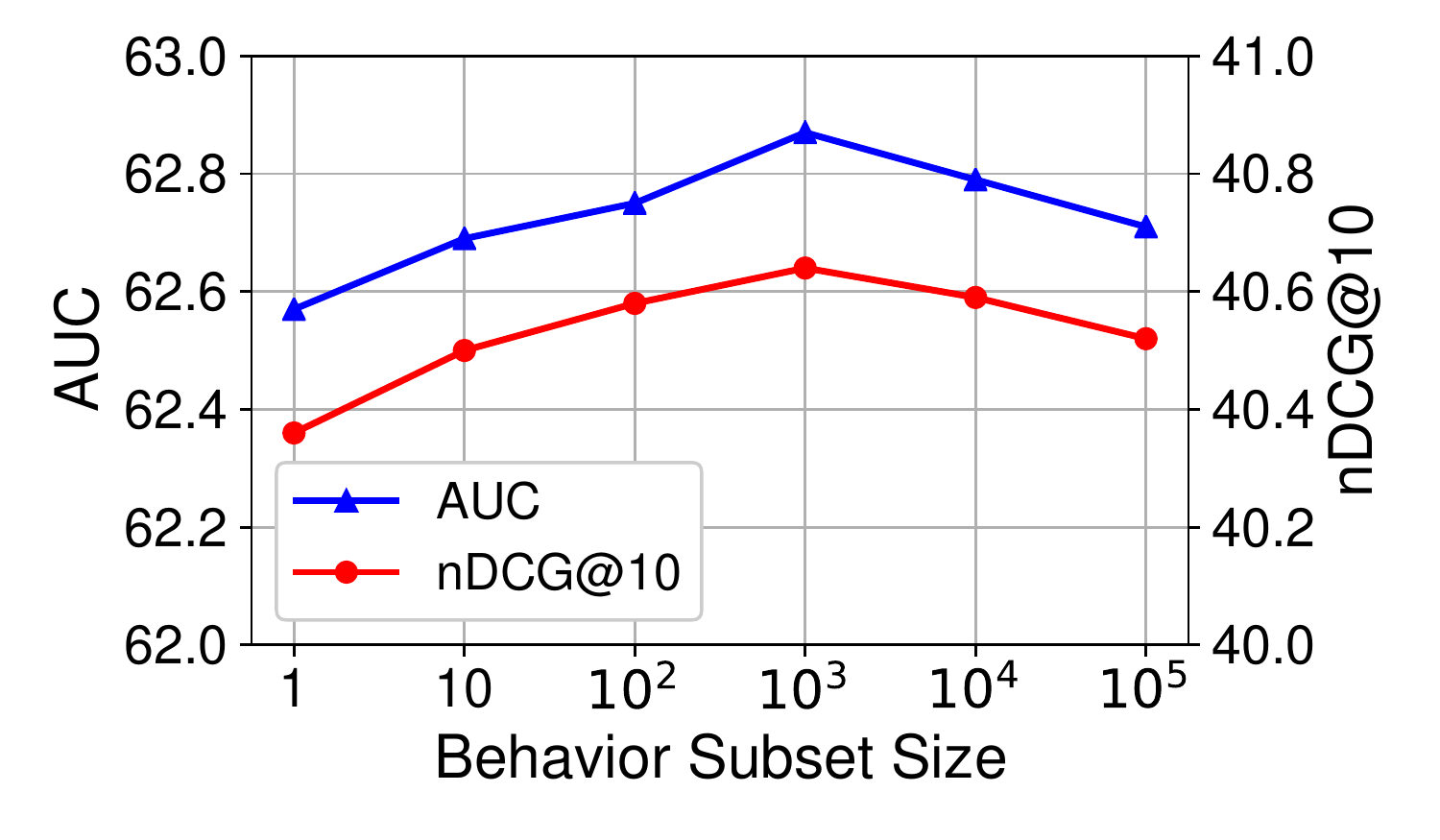}
	\vspace{0.0in}
\caption{Influence of the candidate behavior pool size.}\label{fig.p1}  
\vspace{0.0in}
\end{figure}

\begin{figure}[!t]
	\centering
	\includegraphics[width=0.36\textwidth]{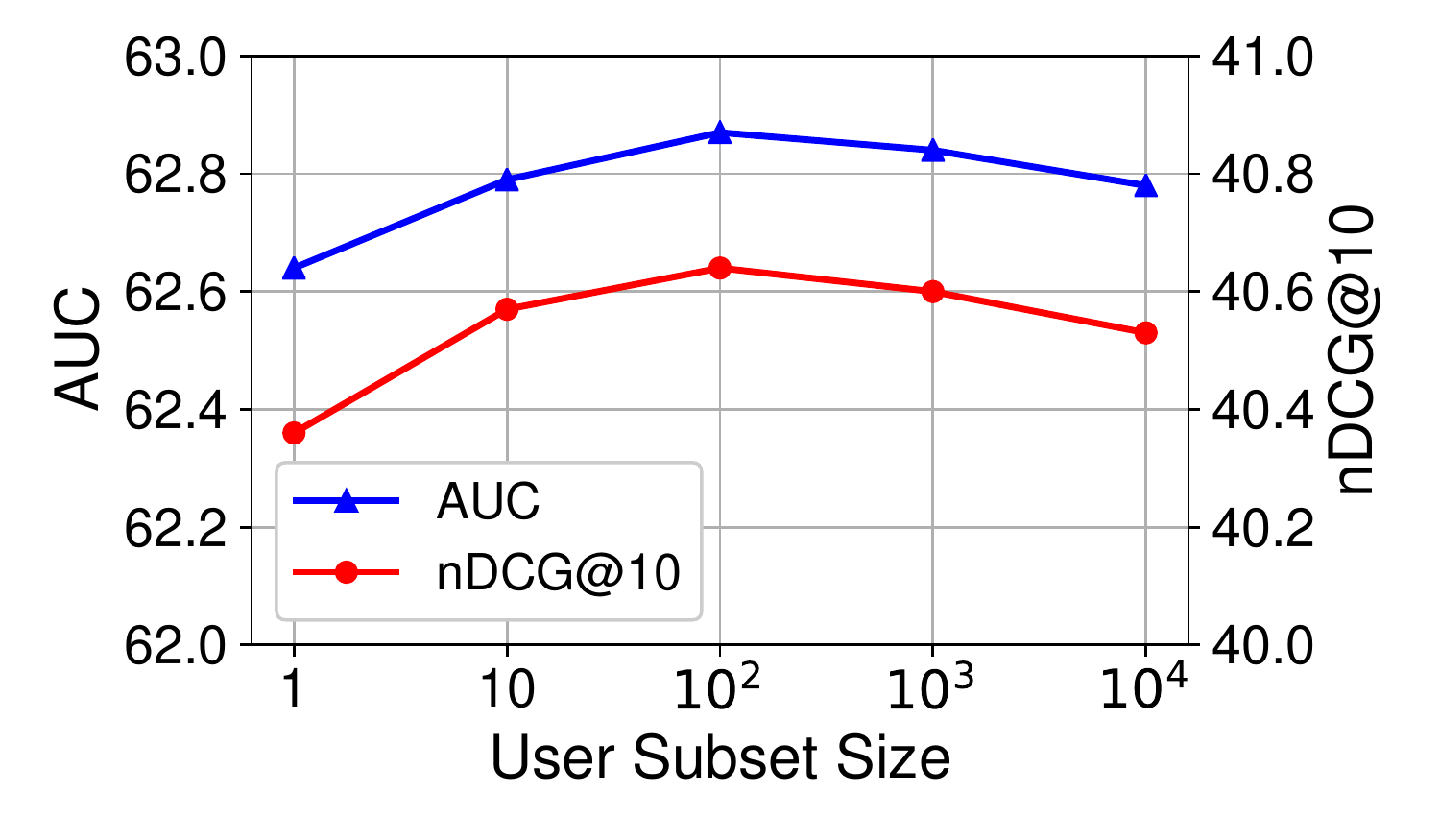}
\caption{Influence of the candidate behavior sequence pool size.}\label{fig.p2} 
\vspace{0.0in} 
\end{figure}

\subsection{Effectiveness of Medium-Hard Negatives}

In this section, we verify the effectiveness of our proposed medium-hard negative sampling framework.
We compare the model performance using random, globally hardest and medium-hard negative candidates, as shown in Fig.~\ref{fig.sample}.
From the results, we find that using globally hardest negatives is slightly better than using random negatives on the \textit{News} dataset.
However, the performance is worse than using random ones on the \textit{CTR} dataset (see supplements).
This may be because the globally hardest negatives may be too difficult for the model to distinguish and may even be misleading.
Our proposed medium-hard negative sampling method outperforms random and hardest negative sampling.
It shows that medium-hard negative candidates  are more suitable for user model pre-training.

\begin{figure}[!t]
	\centering

\subfigure[Model performance.]{\label{fig.ita}
	\includegraphics[width=0.36\textwidth]{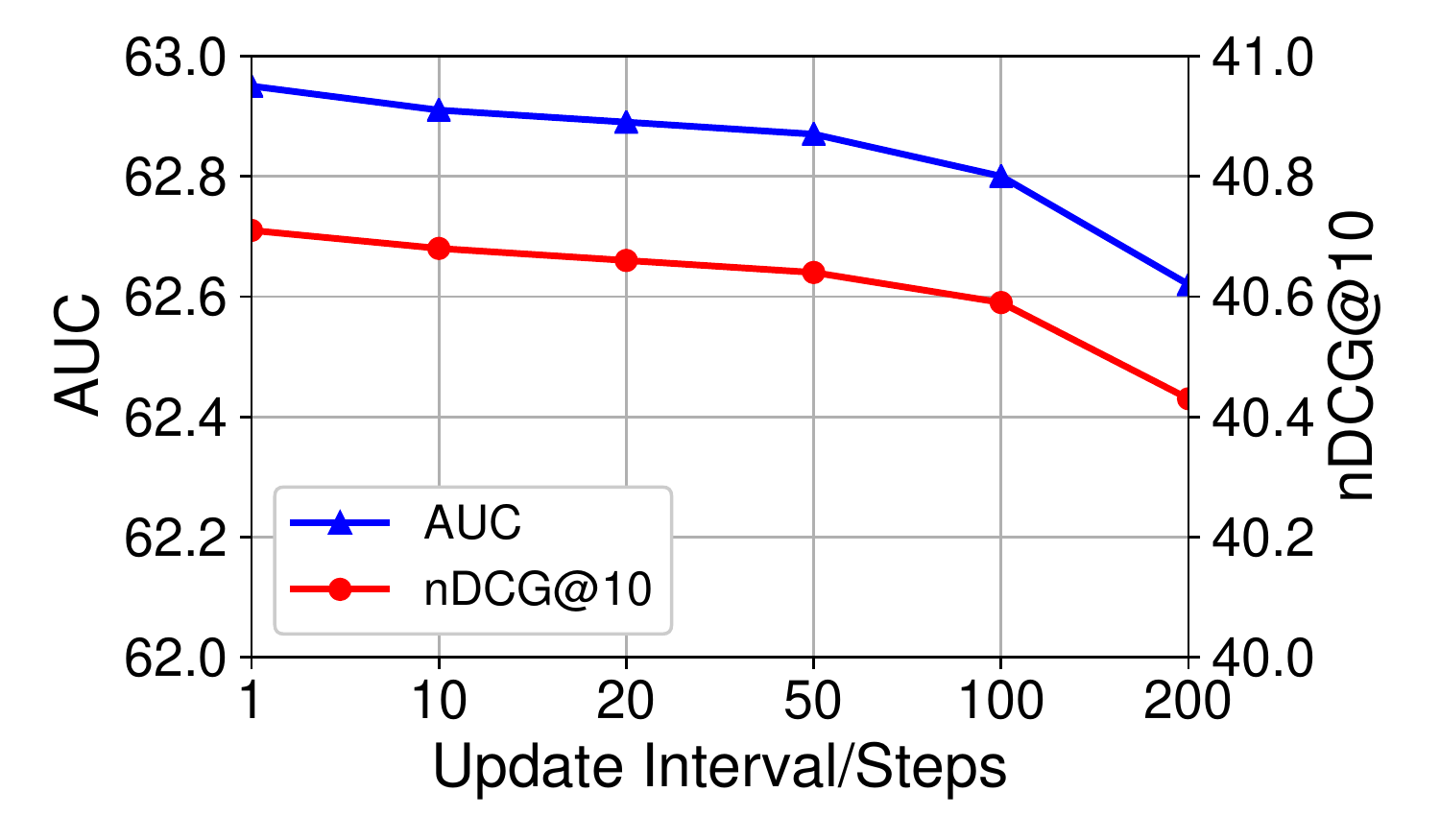}
	}
		\subfigure[Pre-training time.]{\label{fig.itc}
	\includegraphics[width=0.29\textwidth]{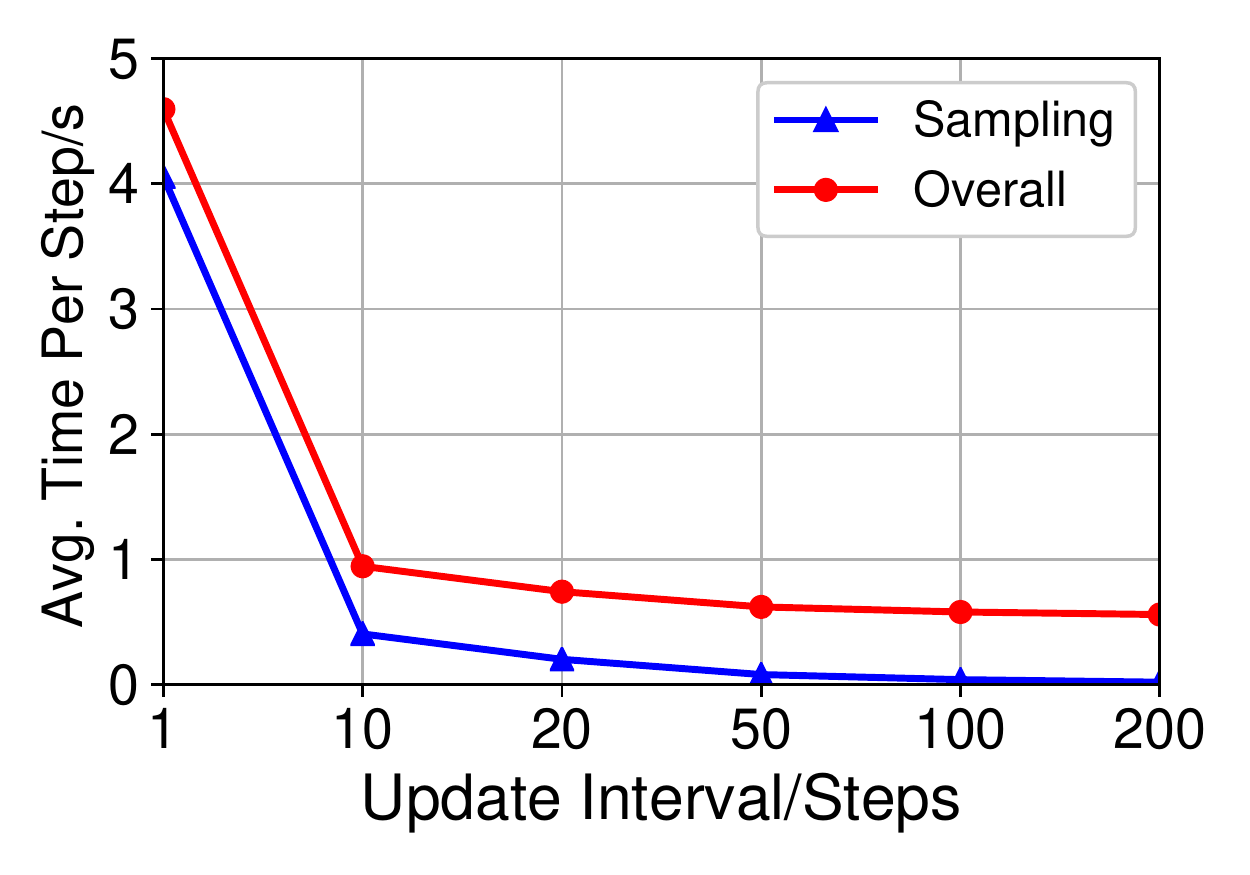}
	}  \vspace{0.0in}
\caption{Influence of the update interval of behavior sequence pool.}\label{fig.p3}
\vspace{0.0in}  
\end{figure}

\subsection{Hyperparameter Analysis}

In this section, we present some analysis on several key hyperparameters in our approach.
We first study the influence of the size of candidate behavior and  behavior sequence pools on the model performance, and the results are respectively shown in Figs.~\ref{fig.p1} and~\ref{fig.p2}.
We find that when the pool sizes are too small, the performance is not optimal because the hardest negative samples in a small candidate pool may still  be easy samples.
However, when the pool sizes get too large, the performance also declines. 
This is because the selected samples will get close to the globally hardest samples, which are too difficult to be distinguished.
Thus, we choose moderate values for the two pool sizes, i.e., 1,000 for the candidate behavior pool and 100 for the candidate behavior sequence pool.

Then, we study the influence of the update interval of the candidate behavior sequence pool.
The model performance, as well as the sampling and overall pre-training time per step under different update intervals are shown in Fig.~\ref{fig.p3}.
We find that a smaller update interval usually yields slightly better performance, but the sampling computational cost is also larger.
The improvement is quite marginal when the interval is smaller than 50 steps while the computational cost grows rapidly.
Thus, we set the update interval to 50 steps to achieve a good trade-off between performance and efficiency.

\section{Conclusion}\label{sec:Conclusion}

In this paper, we propose a UserBERT approach that contrastively pre-trains user models in two  self-supervision tasks, i.e., a masked behavior prediction task to capture relations between user behaviors and a behavior sequence matching task to capture the inherent user interests.
In addition, we propose a medium-hard negative sampling framework to select informative negative candidates to learn accurate and discriminative user models.
Extensive experiments on two real-world datasets show that UserBERT can consistently improve many user models and outperform several baseline methods.


\bibliography{emnlp2021}

\begin{thebibliography}{27}
\expandafter\ifx\csname natexlab\endcsname\relax\def\natexlab#1{#1}\fi

\bibitem[{An et~al.(2019{\natexlab{a}})An, Wu, Wang, Di, Huang, and
  Xie}]{an2019native}
Mingxiao An, Fangzhao Wu, Heyuan Wang, Tao Di, Jianqiang Huang, and Xing Xie.
  2019{\natexlab{a}}.
\newblock Neural ctr prediction for native ad.
\newblock In \emph{CCL}, pages 600--612. Springer.

\bibitem[{An et~al.(2019{\natexlab{b}})An, Wu, Wu, Zhang, Liu, and
  Xie}]{an2019neural}
Mingxiao An, Fangzhao Wu, Chuhan Wu, Kun Zhang, Zheng Liu, and Xing Xie.
  2019{\natexlab{b}}.
\newblock Neural news recommendation with long-and short-term user
  representations.
\newblock In \emph{ACL}, pages 336--345.

\bibitem[{Chen et~al.(2020)Chen, Kornblith, Norouzi, and
  Hinton}]{chen2020simple}
Ting Chen, Simon Kornblith, Mohammad Norouzi, and Geoffrey Hinton. 2020.
\newblock A simple framework for contrastive learning of visual
  representations.
\newblock In \emph{ICML}, pages 1597--1607. PMLR.

\bibitem[{Chi et~al.(2021)Chi, Dong, Wei, Yang, Singhal, Wang, Song, Mao,
  Huang, and Zhou}]{chi2020infoxlm}
Zewen Chi, Li~Dong, Furu Wei, Nan Yang, Saksham Singhal, Wenhui Wang, Xia Song,
  Xian-Ling Mao, Heyan Huang, and Ming Zhou. 2021.
\newblock Infoxlm: An information-theoretic framework for cross-lingual
  language model pre-training.
\newblock In \emph{NAACL-HLT}.

\bibitem[{Devlin et~al.(2019)Devlin, Chang, Lee, and
  Toutanova}]{devlin2019bert}
Jacob Devlin, Ming-Wei Chang, Kenton Lee, and Kristina Toutanova. 2019.
\newblock Bert: Pre-training of deep bidirectional transformers for language
  understanding.
\newblock In \emph{NAACL-HLT}, pages 4171--4186.

\bibitem[{He et~al.(2020)He, Fan, Wu, Xie, and Girshick}]{he2020momentum}
Kaiming He, Haoqi Fan, Yuxin Wu, Saining Xie, and Ross Girshick. 2020.
\newblock Momentum contrast for unsupervised visual representation learning.
\newblock In \emph{CVPR}, pages 9729--9738.

\bibitem[{Hidasi et~al.(2016)Hidasi, Karatzoglou, Baltrunas, and
  Tikk}]{hidasi2016gru}
Bal{\'{a}}zs Hidasi, Alexandros Karatzoglou, Linas Baltrunas, and Domonkos
  Tikk. 2016.
\newblock Session-based recommendations with recurrent neural networks.
\newblock In \emph{ICLR}.

\bibitem[{Kalantidis et~al.(2020)Kalantidis, Sariyildiz, Pion, Weinzaepfel, and
  Larlus}]{kalantidis2020hard}
Yannis Kalantidis, Mert~Bulent Sariyildiz, Noe Pion, Philippe Weinzaepfel, and
  Diane Larlus. 2020.
\newblock Hard negative mixing for contrastive learning.
\newblock \emph{arXiv preprint arXiv:2010.01028}.

\bibitem[{Kim(2014)}]{kim2014convolutional}
Yoon Kim. 2014.
\newblock Convolutional neural networks for sentence classification.
\newblock In \emph{EMNLP}, pages 1746--1751.

\bibitem[{Kingma and Ba(2015)}]{kingma2014adam}
Diederik~P. Kingma and Jimmy Ba. 2015.
\newblock Adam: A method for stochastic optimization.
\newblock In \emph{ICLR}.

\bibitem[{Okura et~al.(2017)Okura, Tagami, Ono, and
  Tajima}]{okura2017embedding}
Shumpei Okura, Yukihiro Tagami, Shingo Ono, and Akira Tajima. 2017.
\newblock Embedding-based news recommendation for millions of users.
\newblock In \emph{KDD}, pages 1933--1942.

\bibitem[{Oord et~al.(2018)Oord, Li, and Vinyals}]{oord2018representation}
Aaron van~den Oord, Yazhe Li, and Oriol Vinyals. 2018.
\newblock Representation learning with contrastive predictive coding.
\newblock \emph{arXiv preprint arXiv:1807.03748}.

\bibitem[{Robinson et~al.(2020)Robinson, Chuang, Sra, and
  Jegelka}]{robinson2020contrastive}
Joshua Robinson, Ching-Yao Chuang, Suvrit Sra, and Stefanie Jegelka. 2020.
\newblock Contrastive learning with hard negative samples.
\newblock \emph{arXiv preprint arXiv:2010.04592}.

\bibitem[{Sun et~al.(2019)Sun, Liu, Wu, Pei, Lin, Ou, and
  Jiang}]{sun2019bert4rec}
Fei Sun, Jun Liu, Jian Wu, Changhua Pei, Xiao Lin, Wenwu Ou, and Peng Jiang.
  2019.
\newblock Bert4rec: Sequential recommendation with bidirectional encoder
  representations from transformer.
\newblock In \emph{CIKM}, pages 1441--1450.

\bibitem[{Vaswani et~al.(2017)Vaswani, Shazeer, Parmar, Uszkoreit, Jones,
  Gomez, Kaiser, and Polosukhin}]{vaswani2017attention}
Ashish Vaswani, Noam Shazeer, Niki Parmar, Jakob Uszkoreit, Llion Jones,
  Aidan~N Gomez, {\L}ukasz Kaiser, and Illia Polosukhin. 2017.
\newblock Attention is all you need.
\newblock In \emph{NIPS}, pages 5998--6008.

\bibitem[{Wu et~al.(2019{\natexlab{a}})Wu, Wu, An, Huang, Huang, and
  Xie}]{wu2019}
Chuhan Wu, Fangzhao Wu, Mingxiao An, Jianqiang Huang, Yongfeng Huang, and Xing
  Xie. 2019{\natexlab{a}}.
\newblock Neural news recommendation with attentive multi-view learning.
\newblock In \emph{IJCAI}, pages 3863--3869.

\bibitem[{Wu et~al.(2019{\natexlab{b}})Wu, Wu, An, Qi, Huang, Huang, and
  Xie}]{wu2019nrhub}
Chuhan Wu, Fangzhao Wu, Mingxiao An, Tao Qi, Jianqiang Huang, Yongfeng Huang,
  and Xing Xie. 2019{\natexlab{b}}.
\newblock Neural news recommendation with heterogeneous user behavior.
\newblock In \emph{EMNLP-IJCNLP}, pages 4876--4885.

\bibitem[{Wu et~al.(2019{\natexlab{c}})Wu, Wu, Ge, Qi, Huang, and
  Xie}]{wu2019nrms}
Chuhan Wu, Fangzhao Wu, Suyu Ge, Tao Qi, Yongfeng Huang, and Xing Xie.
  2019{\natexlab{c}}.
\newblock Neural news recommendation with multi-head self-attention.
\newblock In \emph{EMNLP-IJCNLP}, pages 6390--6395.

\bibitem[{Wu et~al.(2019{\natexlab{d}})Wu, Wu, Liu, He, Huang, and
  Xie}]{wu2019neural}
Chuhan Wu, Fangzhao Wu, Junxin Liu, Shaojian He, Yongfeng Huang, and Xing Xie.
  2019{\natexlab{d}}.
\newblock Neural demographic prediction using search query.
\newblock In \emph{WSDM}, pages 654--662.

\bibitem[{Wu et~al.(2020{\natexlab{a}})Wu, Wu, Qi, Lian, Huang, and
  Xie}]{wu2020ptum}
Chuhan Wu, Fangzhao Wu, Tao Qi, Jianxun Lian, Yongfeng Huang, and Xing Xie.
  2020{\natexlab{a}}.
\newblock Ptum: Pre-training user model from unlabeled user behaviors via
  self-supervision.
\newblock In \emph{EMNLP: Findings}, pages 1939--1944.

\bibitem[{Wu et~al.(2020{\natexlab{b}})Wu, Qiao, Chen, Wu, Qi, Lian, Liu, Xie,
  Gao, Wu et~al.}]{wu2020mind}
Fangzhao Wu, Ying Qiao, Jiun-Hung Chen, Chuhan Wu, Tao Qi, Jianxun Lian,
  Danyang Liu, Xing Xie, Jianfeng Gao, Winnie Wu, et~al. 2020{\natexlab{b}}.
\newblock Mind: A large-scale dataset for news recommendation.
\newblock In \emph{ACL}, pages 3597--3606.

\bibitem[{Xiao et~al.(2017)Xiao, Li, Wang, Lin, and Wang}]{xiao2017joint}
Tong Xiao, Shuang Li, Bochao Wang, Liang Lin, and Xiaogang Wang. 2017.
\newblock Joint detection and identification feature learning for person
  search.
\newblock In \emph{CVPR}, pages 3415--3424.

\bibitem[{Xie et~al.(2020)Xie, Sun, Liu, Gao, Ding, and
  Cui}]{xie2020contrastive}
Xu~Xie, Fei Sun, Zhaoyang Liu, Jinyang Gao, Bolin Ding, and Bin Cui. 2020.
\newblock Contrastive pre-training for sequential recommendation.
\newblock \emph{arXiv preprint arXiv:2010.14395}.

\bibitem[{Xiong et~al.(2020)Xiong, Xiong, Li, Tang, Liu, Bennett, Ahmed, and
  Overwijk}]{xiong2020approximate}
Lee Xiong, Chenyan Xiong, Ye~Li, Kwok-Fung Tang, Jialin Liu, Paul Bennett,
  Junaid Ahmed, and Arnold Overwijk. 2020.
\newblock Approximate nearest neighbor negative contrastive learning for dense
  text retrieval.
\newblock \emph{arXiv preprint arXiv:2007.00808}.

\bibitem[{Yang et~al.(2016)Yang, Yang, Dyer, He, Smola, and
  Hovy}]{yang2016hierarchical}
Zichao Yang, Diyi Yang, Chris Dyer, Xiaodong He, Alex Smola, and Eduard Hovy.
  2016.
\newblock Hierarchical attention networks for document classification.
\newblock In \emph{NAACL-HLT}, pages 1480--1489.

\bibitem[{Yuan et~al.(2020)Yuan, He, Karatzoglou, and
  Zhang}]{yuan2020parameter}
Fajie Yuan, Xiangnan He, Alexandros Karatzoglou, and Liguang Zhang. 2020.
\newblock Parameter-efficient transfer from sequential behaviors for user
  modeling and recommendation.
\newblock In \emph{SIGIR}, pages 1469--1478.

\bibitem[{Zhou et~al.(2019)Zhou, Mou, Fan, Pi, Bian, Zhou, Zhu, and
  Gai}]{zhou2019deep}
Guorui Zhou, Na~Mou, Ying Fan, Qi~Pi, Weijie Bian, Chang Zhou, Xiaoqiang Zhu,
  and Kun Gai. 2019.
\newblock Deep interest evolution network for click-through rate prediction.
\newblock In \emph{AAAI}, volume~33, pages 5941--5948.

\end{thebibliography}
\bibliographystyle{acl_natbib}

\clearpage
\appendix
\section{Appendix}

\subsection{Experiment Environment}

All our experiments are conducted on a Linux server installed with Ubuntu 16.04 operating system and Python 3.7.
The CPU type is Intel Xeon E5-2620 v4, and the type of GPU is GeForce GTX1080Ti with 12 GB memory.
The total memory is 64GB.
We use Keras 2.2.4 with the tensorflow 1.12 backend to implement our models.
Each experiment is run on a single GPU and CPU core with a single process/thread.

\subsection{Preprocessing}
In our approach, we the the word tokenizer in NLTK and we use at most 30 words in each webpage title.
In addition, we only keep at most 100 behaviors of each user.
All behavior texts are filled with zero padding tokens and are padded to the same length.
The user behavior sequence  is padded with empty user behaviors.

\subsection{Hyperparameter Settings}

The detailed hyperparameter settings are listed in Table~\ref{hyper}.

\begin{table}[h]
\centering
\resizebox{1.0\linewidth}{!}{
\begin{tabular}{|l|c|c|c|}
\hline
\multicolumn{1}{|c|}{\textbf{Hyperparameters}}& \textit{Pretrain} & \textit{News} & \textit{CTR}\\ \hline
hidden dimension                     & 256   & 256  & 256           \\ 
CNN  window size                     & 3   & 3   & 3             \\ 
attention query dimension                     & 200 & 200 & 200              \\ 
behavior pool size                                   & 1,000 & - & -      \\ 
behavior sequence pool size                                   & 100 & - & -      \\ 
update interval                                   & 50 & - & -      \\ 
negative sampling ratio                     & 4 & 4 & 1              \\ 
dropout                                      & 0.2 & 0.2 &0.2              \\
optimizer                                    & Adam & Adam & Adam             \\
learning rate                                 & 1e-5         & 1e-4 & 1e-4      \\ 
batch size                                   & 32 & 32 & 64      \\   
\hline
\end{tabular}
}
\caption{Detailed hyperparameter settings.}\label{hyper}
\end{table}

\subsection{Experiments on the CTR Dataset}

\begin{figure}[h]
	\centering

	\includegraphics[width=0.38\textwidth]{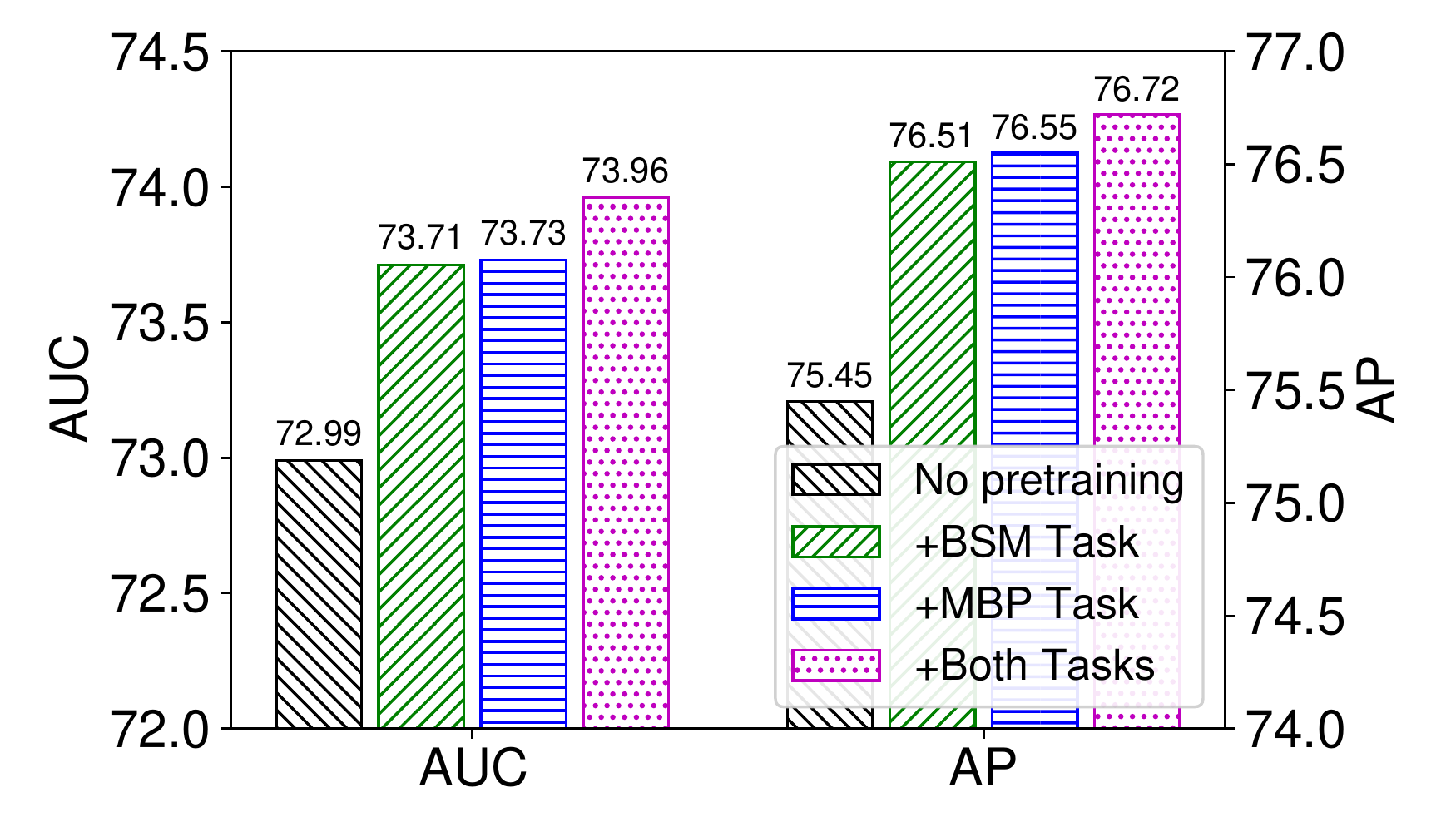}
\vspace{0.0in}
\caption{Effect of the two pre-training tasks.}
	  \label{fig.taskb}
\vspace{0.0in}
\end{figure}

\begin{figure}[h]
	\centering
	\includegraphics[width=0.38\textwidth]{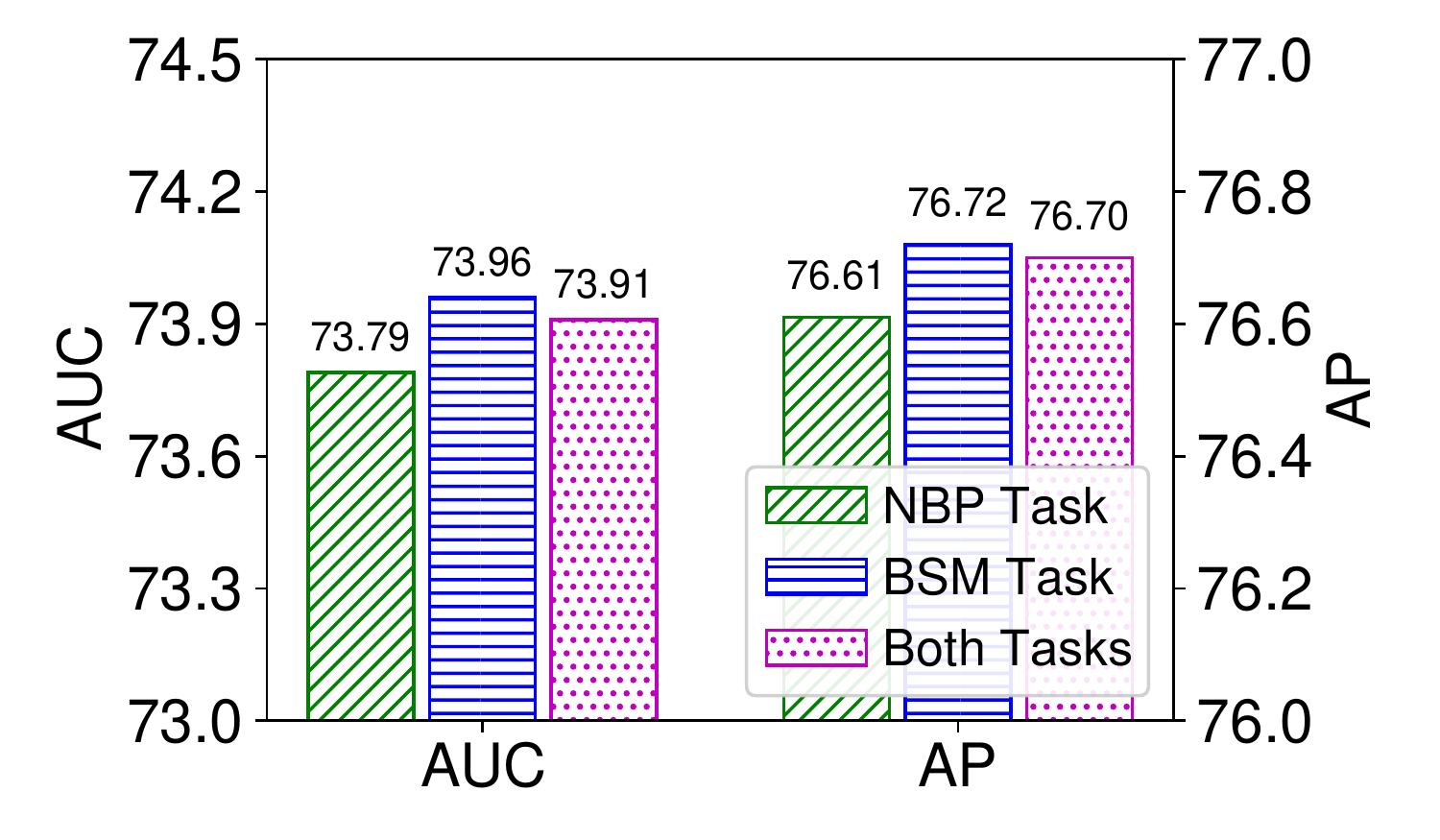}
	\vspace{0.0in}
\caption{Comparisons between the BSM task in UserBERT and the NBP task in PTUM.}\label{fig.taskb2} 
\vspace{0.0in} 
\end{figure}

\begin{figure}[h]
	\centering
	\includegraphics[width=0.38\textwidth]{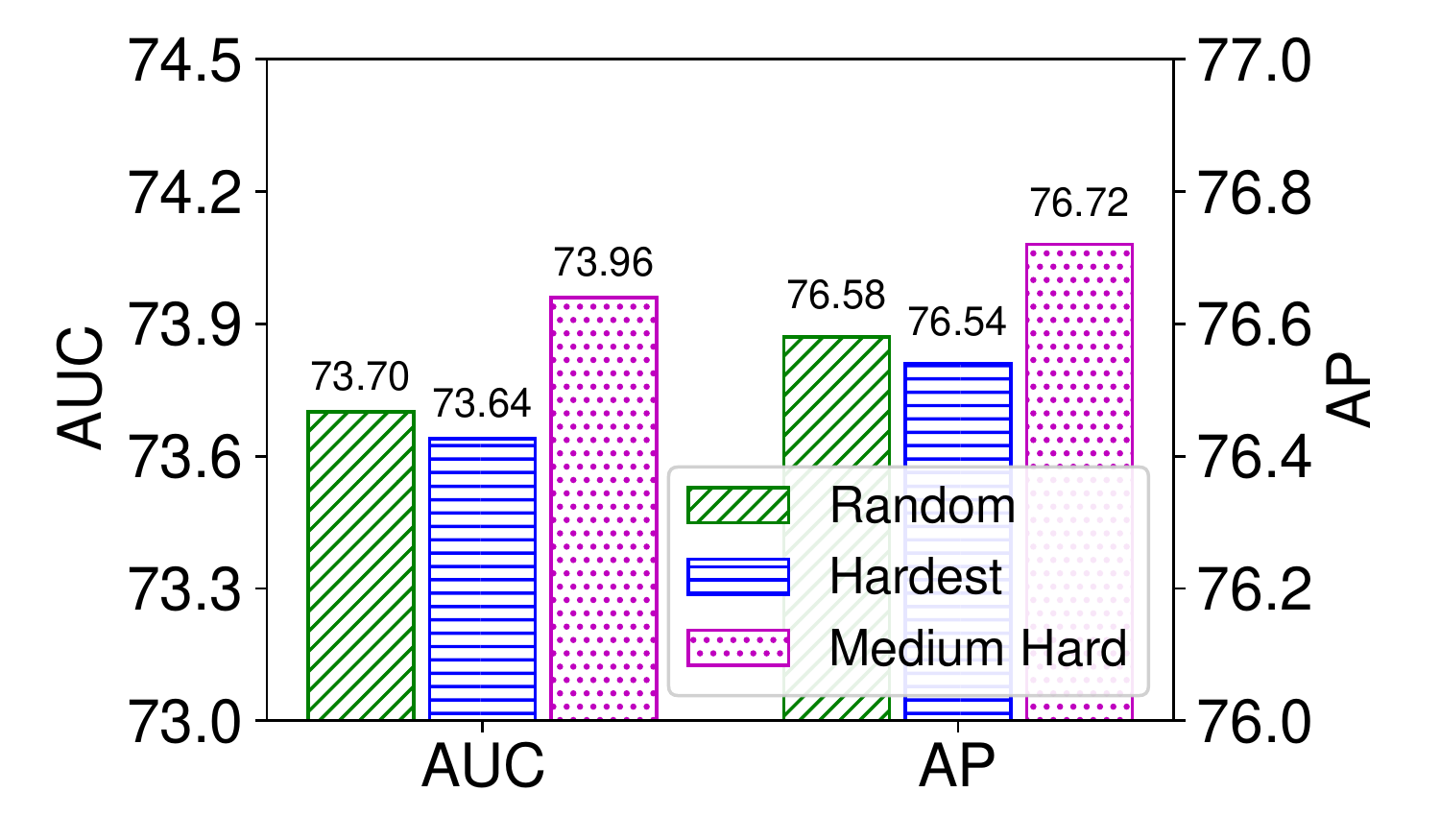}
	
	\vspace{0.0in}
\caption{Effect of medium-hard negative sampling.}\label{fig.sampleb}  
\vspace{0.0in}
\end{figure}

\begin{figure}[h]
	\centering

	\includegraphics[width=0.38\textwidth]{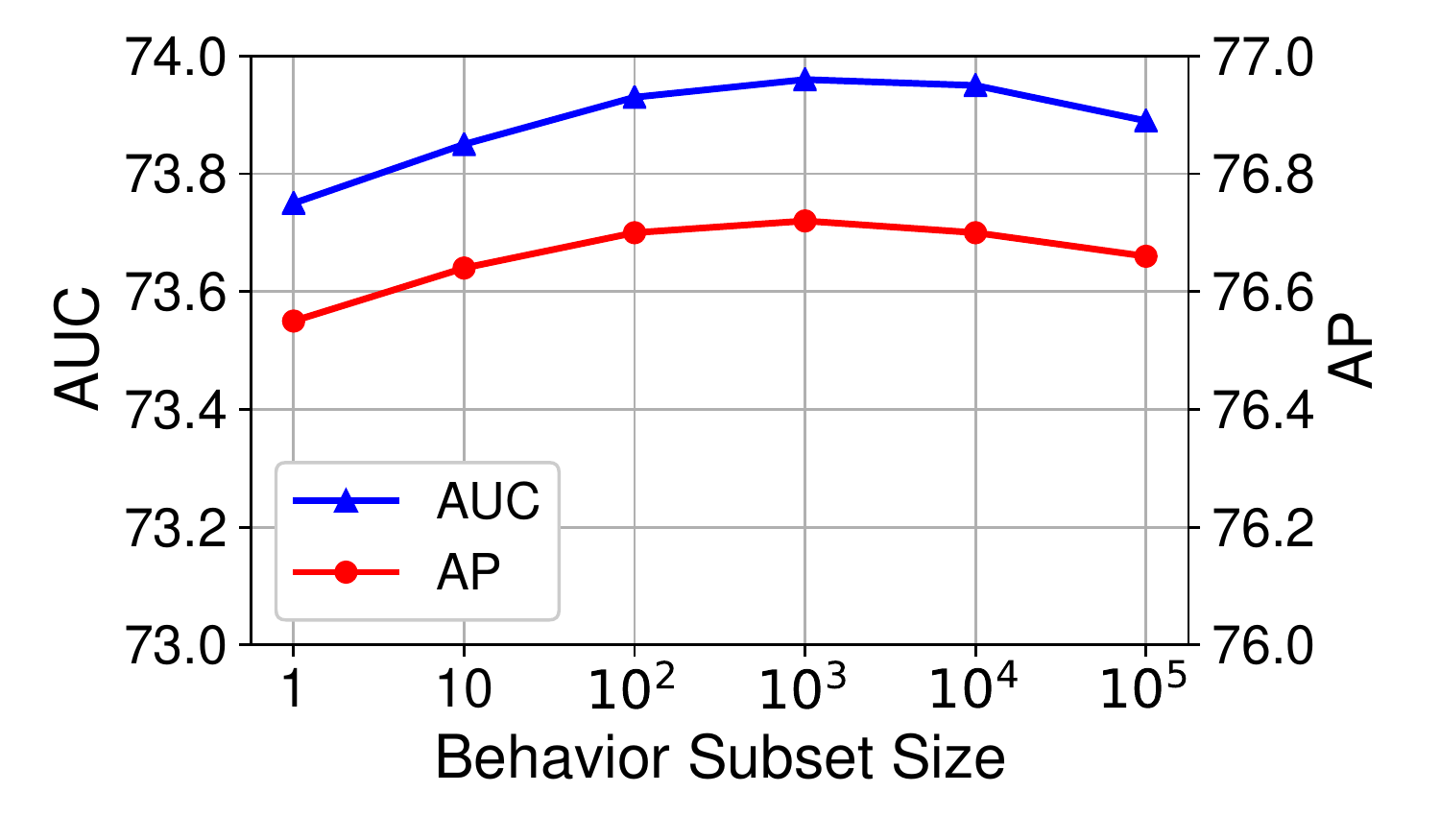}
	
	\vspace{0.0in}
\caption{Influence of the candidate behavior pool size.}\label{fig.pb1}  
\vspace{0.0in}
\end{figure}

\begin{figure}[h]
	\centering
	\includegraphics[width=0.38\textwidth]{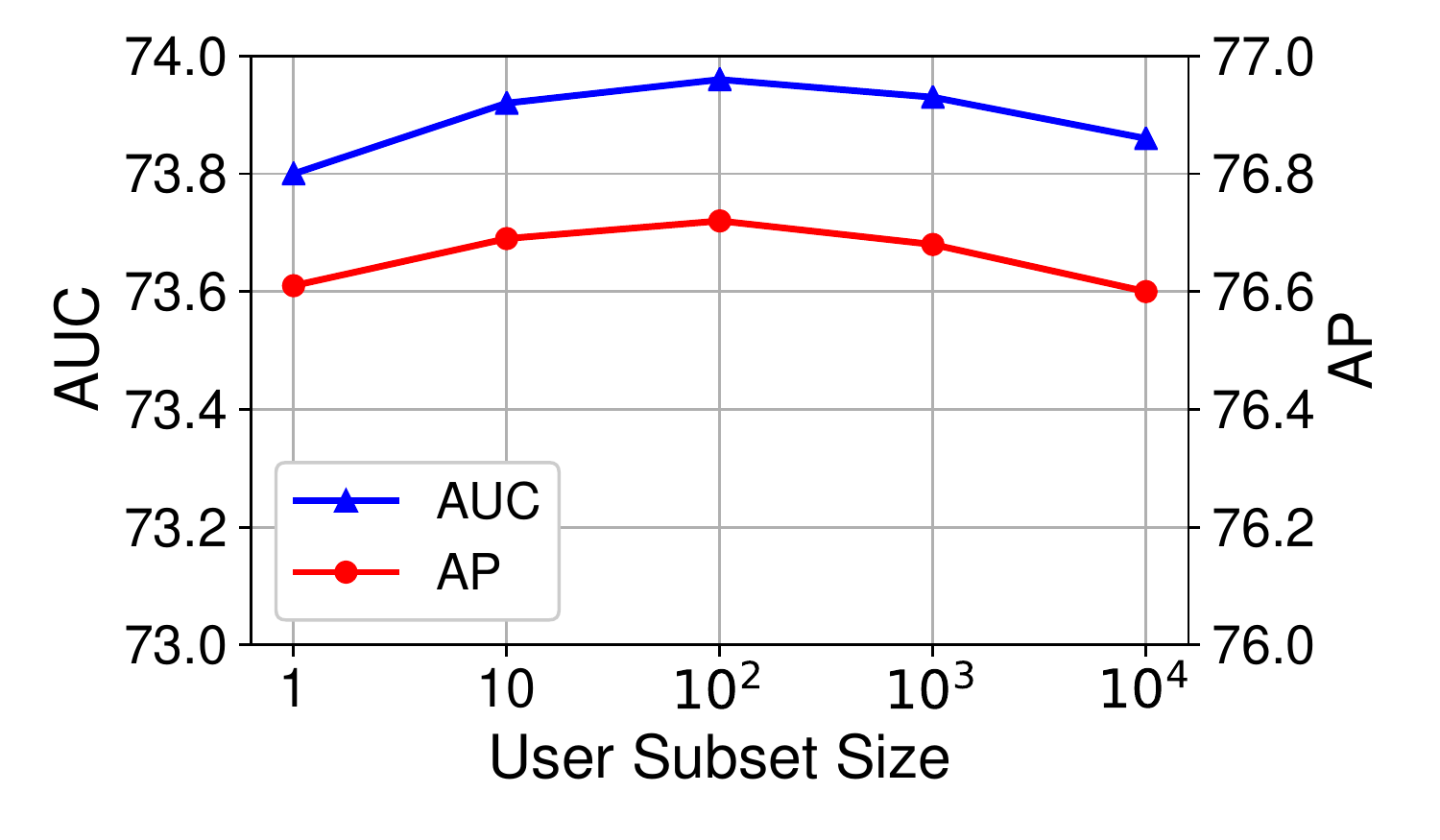}
 \vspace{0.0in}
\caption{Influence of the candidate behavior sequence pool size.}\label{fig.pb2} 
\vspace{0.0in} 
\end{figure}

\begin{figure}[h]
	\centering	\includegraphics[width=0.38\textwidth]{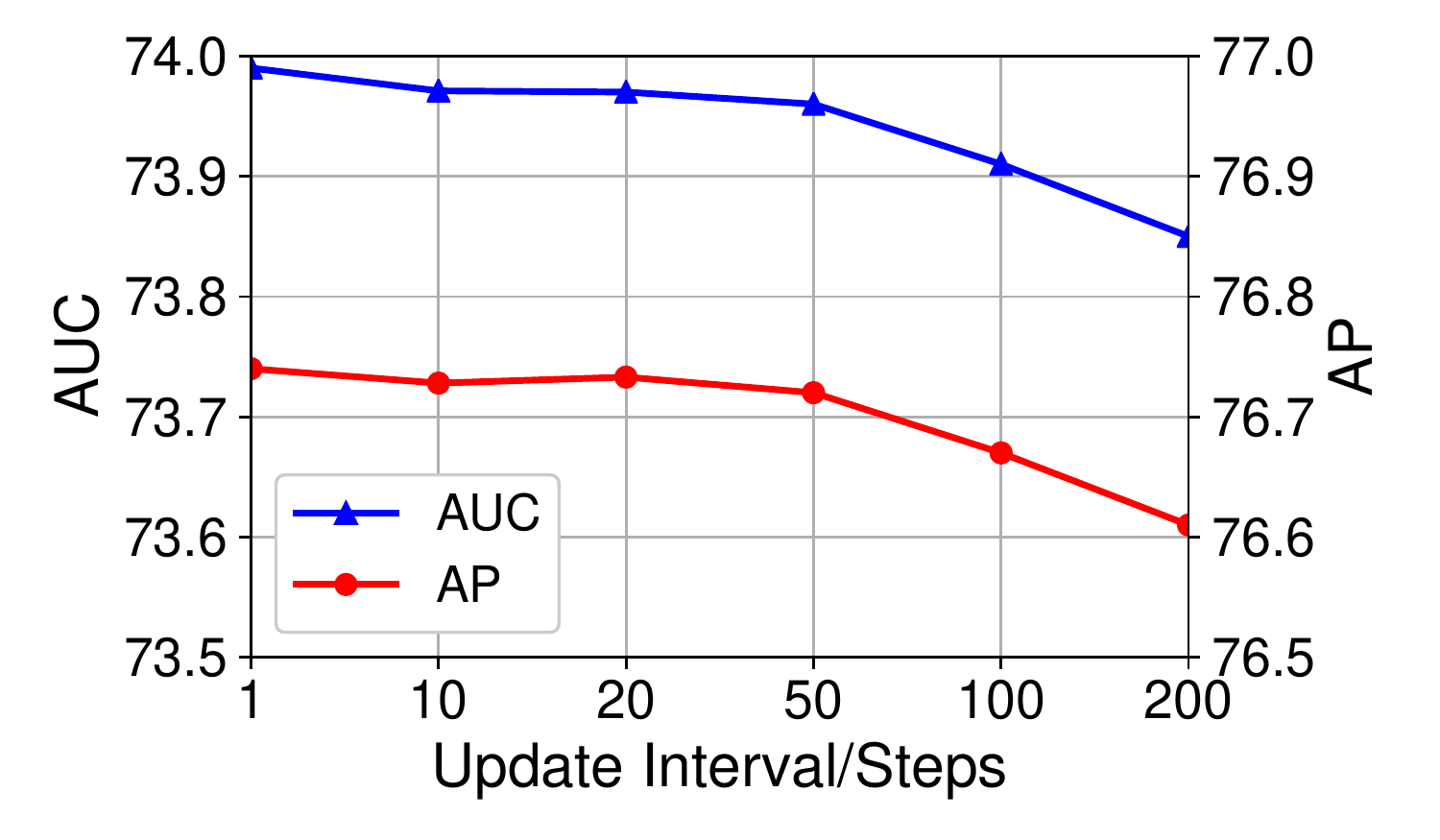}
 \vspace{0.0in}
\caption{Influence of the update interval of behavior sequence pool on the model performance.}\label{fig.itb}
\vspace{0.0in}  
\end{figure}

The experimental results on the \textit{CTR} dataset  are shown in Figs.~\ref{fig.taskb}-\ref{fig.itb}.

\end{document}